\documentclass[preprint,11pt,nofootinbib]{revtex4-1}
	\usepackage{amssymb}
	\usepackage{amsmath}
	\usepackage{hyperref}
	\usepackage{graphicx}
	\usepackage{color}
	\newcommand{\nn}{\nonumber}
	\newcommand{\lam}{\lambda}
	\newcommand{\rbra}[1]{\left(#1\right)}
	\newcommand{\sbra}[1]{\left[#1\right]}
	\newcommand{\cbra}[1]{\left\{#1\right\}}
	\newcommand{\eval}[1]{\langle#1\rangle}
	\newcommand{\re}{\textrm{Re}}
	\newcommand{\im}{\textrm{Im}}
	\newcommand{\eq}[1]{\begin{align}#1\end{align}}
	\newcommand{\texteq}[1]{$#1$}
	\newcommand{\primed}[1]{{#1'}}

\allowdisplaybreaks[4]

\begin{document}

\title{Testing aligned CP-violating Higgs sector at future lepton colliders}

%%%%%     authors     %%%%%
\author{Shinya Kanemura}
\email{kanemu@het.phys.sci.osaka-u.ac.jp}
\author{Mitsunori Kubota}
\email{mkubota@het.phys.sci.osaka-u.ac.jp}
\author{Kei Yagyu}
\email{yagyu@het.phys.sci.osaka-u.ac.jp}
\affiliation{Department of Physics, Osaka University, Toyonaka, Osaka 560-0043, Japan}

%%%%%     date     %%%%%
%\date{\today}

%%%%%     abstract     %%%%%
\begin{abstract}
We discuss the testability of CP-violating phases at future lepton colliders for the scenario which satisfies electric dipole moment data by destructive interferences among several phases. 
We consider the general but aligned two Higgs doublet model which has the CP-violating phases in the Higgs potential and the Yukawa interaction. 
The Yukawa interaction terms are aligned to avoid flavor changing neutral currents at tree level. 
The Higgs potential is also aligned such that the coupling constants of the lightest Higgs boson with the mass of 125 GeV to the Standard Model (SM) particles are the same as those of the SM at tree level. 
We investigate the azimuthal angle distribution of the hadronic decay of tau leptons arising from production and decay of the extra Higgs bosons, which contains information of the CP-violating phases. 
From the signal and background simulation, we find that the scenario with finite CP-violating phases can be distinguished from CP conserving one at future lepton colliders like the International Linear Collider. 
\end{abstract}

%%%%%     preprint     %%%%%
\preprint{OU-HET-1084}

%%%%%%%%%%
%%%%%%%%%%
\maketitle

%%%%%     table of contents     %%%%%
%\tableofcontents
%\newpage

%%%%%%%%%%     introduction     %%%%%%%%%%
\section{Introduction}
Baryon Asymmetry of the Universe (BAU) is one of the fundamental questions of our universe, which cannot be explained in the Standard Model (SM) for particle physics. 
Baryogenesis is the most reasonable idea to explain BAU. 
There are various scenarios for the realization of baryogenesis, such as GUT baryogenesis\cite{Yoshimura:1978ex,Weinberg:1979bt}, leptogenesis\cite{Fukugita:1986hr}, electroweak baryogenesis\cite{Kuzmin:1985mm} and so on. 

Electroweak baryogenesis is a promising scenario which relies on the structure of the sector of electroweak symmetry breaking, in which the Sakharov conditions\cite{Sakharov:1967dj} are satisfied by the sphaleron transition, additional CP violation in the extension of the Higgs sector and the strongly first order electroweak phase transition. 
These can be realized by introducing an extended Higgs sector, and various models of electroweak baryogenesis have been investigated along this line such as those with additional isospin singlets\cite{Espinosa:2011eu,Cline:2012hg,Grzadkowski:2018nbc}, doublets\cite{Turok:1990zg,Cline:1995dg,Fromme:2006cm,Cline:2011mm,Shu:2013uua,Fuyuto:2017ewj}, triplets\cite{Patel:2012pi,Chiang:2014hia} and so on. 
One of the simplest and interesting candidates is the two Higgs doublet model (THDM)\cite{Lee:1973iz}, in which additional CP-violating phases can be provided by the Higgs potential and the Yukawa interactions. 
In addition, the first order phase transition can be easily realized by the mixing among the scalar bosons\cite{Pietroni:1992in}, or by the quantum non-decoupling loop effects on the effective potential due to additional scalar fields in this model\cite{Kanemura:2004ch}. 

The parameter space of the THDM has been constrained precision measurements at LEP/SLC experiments, LHC data\cite{Celis:2013rcs,Chowdhury:2017aav,Aiko:2020ksl,Eberhardt:2020dat} and various flavor experiments\cite{Mahmoudi:2009zx,Jung:2010ab}. 
Nevertheless, there is still a wide region of the parameter space for the scenario of electroweak baryogenesis\cite{Dorsch:2016nrg,Fuyuto:2019svr}. 

There are notable predictions in part of scenarios, where the first order phase transition is realized by the non-decoupling loop effect. 
Such a non-decoupling effect introduced to realize the strongly first order phase transition can also affect various Higgs observables such as the di-photon decay\cite{Ginzburg:2001ph,Gil:2012ya,Hashino:2015nxa} of the Higgs boson and the triple Higgs boson coupling\cite{Kanemura:2002vm,Kanemura:2004mg,Braathen:2019pxr,Braathen:2019zoh}. 
This important features can be tested at future collider experiments such as the high-luminosity upgrade of the LHC (HL-LHC)\cite{Cepeda:2019klc} and the International Linear Collider (ILC)\cite{Baer:2013cma,Fujii:2015jha,Fujii:2017vwa}. 
At the same time, the strongly first order phase transition in the early universe produces static gravitational waves with a specific shape of the spectrum\cite{Grojean:2006bp,Espinosa:2007qk,Espinosa:2008kw,Espinosa:2010hh,Hindmarsh:2015qta,Kakizaki:2015wua,Hashino:2016rvx,Hashino:2018wee,Zhou:2020xqi}, which can be also tested at future space-based gravitational wave interferometers such as LISA\cite{Audley:2017drz}, DECIGO \cite{Seto:2001qf} and BBO\cite{Corbin:2005ny}. 

One of the most serious constraints on the CP-violating THDM is those from electric dipole moments (EDMs)\cite{Bernreuther:1990jx,Fukuyama:2012np,Leigh:1990kf,BowserChao:1997bb,Jung:2013hka,Abe:2013qla,Cheung:2014oaa,Cheung:2020ugr,Egana-Ugrinovic:2018fpy,Altmannshofer:2020shb}. 
Current experimental bounds on the electron EDM and the neutron EDM are given in Refs.\cite{Andreev:2018ayy,Abel:2020gbr}. 
It is getting common that it is not easy to build a realistic scenario for successful electroweak baryogenesis with satisfying the EDM data in the THDM. 
Other scenarios such as those with singlet extension are also being explored\cite{Espinosa:2011eu,Cline:2012hg,Grzadkowski:2018nbc}. 

In our recent paper\cite{Kanemura:2020ibp}, however, we have proposed a new scenario of the aligned THDM, in which CP-violating effects from the Higgs potential and the Yukawa interaction destructively interfere on the EDM, so that the current EDM constraints can be avoided. 
In this scenario, two kinds of the alignment are imposed. 
Flavor changing neutral currents (FCNCs) at tree level are avoided by the alignment on the Yukawa interaction terms\cite{Pich:2009sp}. 
The other alignment on the Higgs potential realizes that the coupling constants of the lightest Higgs boson with the mass of 125 GeV to the SM particles are the same as those of the SM at tree level. 
We have shown that a sufficient amount of CP-violating phases can exist in the Higgs potential and the Yukawa interaction, which can in principle reproduce the current abundance of the baryon number. 

In this paper, we investigate to test this CP-violating scenario of the aligned THDM using the precision measurements of the decays of the additional Higgs bosons at future lepton collider experiments. 
In such a CP-violating model with the alignment, the collider phenomenology of the extra Higgs bosons plays an important role to test the CP violation. 
In particular, we focus on the decay of the extra Higgs bosons into a tau lepton pair in order to see effects of the CP-violating phases from their kinematic structures. 
Testing CP violating effects from the tau decay of the 125 GeV Higgs boson have been studied at the LHC in Refs.~\cite{Harnik:2013aja} and at electron positron colliders in Refs.~\cite{Grzadkowski:1995rx,Harnik:2013aja,Jeans:2018anq,Ge:2020mcl}\footnote{The impact of the CP-violating phases in the Higgs sector on the Higgs boson decays may also be tested at future lepton colliders\cite{Aoki:2018zgq}. }. 

We calculate the azimuthal angle distribution of the hadronic decay of tau leptons at the ILC by using MadGraph5\cite{Alwall:2011uj} and TauDecay\cite{Hagiwara:2012vz}. 
From the signal and background analysis, we find that the scenario with finite CP-violating phases, which satisfies the current EDM data, can be distinguished from the CP-conserving scenario at the energy upgraded version of the ILC. 

This paper is organized as follows. 
In Sec.~\ref{sc:model}, we give a brief review of the CP-violating THDM with two alignments for the Higgs potential and for the Yukawa interaction. 
In Sec.~\ref{sc:EDM}, we discuss the constraint of the several EDMs and the destructive interference between the Barr-Zee (BZ) contributions in our model. 
In Sec.~\ref{sc:CPV}, we discuss the decay of the Higgs bosons. 
In Sec.~\ref{sc:collider}, we show the results of the simulation study for the angular distribution of hadronic tau decays from the production and decay of the extra Higgs bosons. 
We summarize our results in Sec.~\ref{sc:summary}. 
In Appendices~\ref{sc:BZformulae} and \ref{sc:decayrate}, we present the analytic formulae of the BZ type contributions to the EDM (chromo EDM) for any fermions (quarks), and those of partial decay widths for the extra Higgs bosons, respectively.

%%%%%%%%%%     model     %%%%%%%%%%
\section{Aligned two Higgs doublet model}\label{sc:model}
We consider the CP-violating THDM in which the scalar sector consists of two isospin doublets. 
We do not impose any symmetries other than the SM gauge symmetry. 
In our model, we assume that the scalar potential $\mathcal{V}$ is aligned to realize the SM-ilke couplings for the 125 GeV Higgs boson, and the Yukawa interaction term $\mathcal{L}_\textrm{yukawa}$ is also aligned to remove new contributions to FCNCs at tree level. 
Then, we show that additional CP-violating phases appear in the Higgs potential and the Yukawa interaction.

%%%%%     potential     %%%%%
\subsection{Higgs Potential}
The most general form of the Higgs potential is given by
	%%%   potential
		\begin{align}
			\mathcal{V}=
			&	-\mu_1^2 (\Phi_1^\dagger\Phi_1)
				-\mu_2^2 (\Phi_2^\dagger\Phi_2)
				-\sbra{ \mu_3^2 (\Phi_1^\dagger\Phi_2)+h.c.}
			\nn\\
			&	+\tfrac{1}{2} \lam_1 (\Phi_1^\dagger\Phi_1)^2
				+\tfrac{1}{2} \lam_2 (\Phi_2^\dagger\Phi_2)^2
				+\lam_3 (\Phi_1^\dagger\Phi_1) (\Phi_2^\dagger\Phi_2)
				+\lam_4 (\Phi_2^\dagger\Phi_1) (\Phi_1^\dagger\Phi_2)
			\nn\\
			&	+\cbra{\sbra{
					\tfrac{1}{2} \lam_5 (\Phi_1^\dagger\Phi_2)
					+\lam_6 (\Phi_1^\dagger\Phi_1)
					+\lam_7 (\Phi_2^\dagger\Phi_2)
				} (\Phi_1^\dagger\Phi_2)+h.c.}
		\label{eq:potential2}
		,\end{align}
where $\mu_{1,2}^2$ and $\lam_{1,2,3,4}$ are real, while $\mu_3^2$ and $\lam_{5,6,7}$ are complex. 
Without loss of generality, the two isodoublet fields $(\Phi_1,\Phi_2)$ are taken to be the Higgs basis\cite{Botella:1994cs,Davidson:2005cw} defined as
	%%%   scalar doublets
		\begin{align}
			\Phi_1=\rbra{\begin{array}{c} G^+\\\frac{1}{\sqrt{2}}(v+h^0_1+iG^0) \end{array}}
			,\quad
			\Phi_2=\rbra{\begin{array}{c} H^+\\\frac{1}{\sqrt{2}}(h^0_2+ih^0_3) \end{array}}
		\label{eq:parametrizeddoublets}
		,\end{align}
where $G^\pm$ and $G^0$ are the Nambu-Goldstone bosons, $H^\pm$ are the charged Higgs bosons and $h^0_j$ ($j=1,2, 3$) are the neutral Higgs bosons. The vacuum expectation value $v$ is related to the Fermi constant $G_F$ by $v\equiv(\sqrt{2}G_F)^{-1/2}$. 
We can move to the general basis of the Higgs doublets by the $U(2)$ transformation. 
The relation between the parameters of the Higgs potential in the different bases can be found in Ref.~\cite{Kanemura:2020ibp}. 
The stationary conditions for the Higgs potential
	%%%   stationary condition
		\begin{align}
			0=	\left. 
					\frac{\partial\mathcal{V}}{\partial h^0_j}
				\right|_{\substack{\Phi_1=\eval{\Phi_1}\\\Phi_2=\eval{\Phi_2}}}
		,\end{align}
lead to
		\begin{align}
			\mu_1^2 = \frac{1}{2}\lam_1v^2
		,~	\mu_3^2 = \frac{1}{2}\lam_6v^2
		.\end{align}
The remaining dimensionful parameter is redefined as $M^2\equiv-\mu_2^2$ below. 
The squared mass of the charged Higgs boson is given by
	%%%   charged higgs mass
		\begin{align}
			m_{H^\pm}^2=M^2+\tfrac{1}{2}\lam_3v^2
		.\end{align}
The squared-mass matrix for the neutral Higgs bosons in the basis of $(h^0_1,h^0_2,h^0_3)$ is given by
	%%%   mass matrix for h^0
		\small
		\begin{align}
			\mathcal{M}^2
				=	v^2
					\rbra{
					\begin{array}{ccc}
						\lam_1	&\re[\lam_6]	&-\im[\lam_6]	\\
						\re[\lam_6]	&\frac{M^2}{v^2}+\frac{1}{2}(\lam_3+\lam_4+\re[\lam_5])	&-\frac{1}{2}\im[\lam_5]	\\
						-\im[\lam_6]	&-\frac{1}{2}\im[\lam_5]	&\frac{M^2}{v^2}+\frac{1}{2}(\lam_3+\lam_4-\re[\lam_5])
					\end{array}
					}
		\label{eq:massmatrix}
		.\end{align}
		\normalsize
This is diagonalized by the orthogonal matrix $\mathcal{R}$ as $\mathcal{R}^T\mathcal{M}^2\mathcal{R}=\mathrm{diag}(m_{H^0_1}^2,m_{H^0_2}^2,m_{H^0_3}^2)$. 
The mass eigenstates of the neutral Higgs bosons are expressed as
		\eq{
			h^0_i=\mathcal{R}_{ij}H^0_j
		.}

In the model, since one of the complex phases can be absorbed by redefinition of the two doublet fields, the Higgs potential has 11 parameters which are $v, M, \lam_{1,2,3,4}, |\lam_{5,6,7}|$ and the 2 physical phases. 
Hereafter, we take $\arg[\lam_5]=0$ by using the phase redefinition,
	%%%   phase redefinition
		\texteq{
			(\Phi_1^\dagger\Phi_2)\to e^{-\arg[\lam_5]/2}(\Phi_1^\dagger\Phi_2)
		,}
and we also redefine the other complex parameters as
		\texteq{
			\lam_{6,7}e^{-\arg[\lam_5]/2}\to\lam_{6,7}
		.}

We assume the following alignment for the Higgs potential,
	%%%   potential alignment
		\eq{
			\lam_6=0
		,}
in which the mixing matrix is diagonalized as $\mathcal{R}_{ij}=\delta_{ij}$, so that the neutral Higgs bosons do not mix with each other. 
The squared masses of the neutral Higgs bosons are then given by
	%%%   mass of neutral scalars
		\eq{
			m_{H^0_1}^2 &= \lam_1 v^2
		,\\	m_{H^0_2}^2 &= M^2+\frac{1}{2}(\lam_3+\lam_4+\re[\lam_5]) v^2
		,\\	m_{H^0_3}^2 &= M^2+\frac{1}{2}(\lam_3+\lam_4-\re[\lam_5]) v^2
		.}
We identify $H^0_1$ as the discovered Higgs boson with the mass of 125 GeV, and we consider that the other Higgs bosons $H^0_{2,3}$ and $H^\pm$ are heavier. 
Consequently, there are 7 free parameters which can be chosen as follows
	%%%   parameters
		\eq{
			M, m_{H^0_2}, m_{H^0_3}, m_{H^\pm}, \lam_2, |\lam_7| ~\textrm{and}~ \theta_7,
		}
where $\theta_7\equiv\arg[\lam_7]\in(-\pi,\pi]$. 

%%%%%     yukawa     %%%%%
\subsection{Yukawa Interaction}
The most general form of Yukawa interactions is given in terms of $\Phi_1$ and $\Phi_2$ as follows 
	%%%   yukawa term
		\begin{align}
			\mathcal{L}_\textrm{yukawa}&=
				-\sum_{k=1}^2
					\rbra{
					\bar{\primed{Q}}_L y_{u,k}^{\dagger} \tilde{\Phi}_k \primed{u}_R
					+\bar{\primed{Q}}_L y_{d,k} \Phi_k \primed{d}_R
					+\bar{\primed{L}}_L y_{e,k} \Phi_k \primed{e}_R
					+h.c. },
		\label{eq:yukawa}
		\end{align}
where $\tilde{{\Phi}}_{k}=i \sigma_2 {\Phi}_k^*$ and $y_{f,k}$ are the $3\times3$ complex Yukawa coupling matrices in the weak basis for the fermions. 
The left-handed quark and lepton doublets are defined as $\primed{Q}_L$ and $\primed{L}_L$, and the right-handed up-type quark, down-type quark and charged lepton singlets are defined as $\primed{u}_R$, $\primed{d}_R$ and $\primed{e}_R$, respetively. 
The mass matrices for up-type quarks, down-type quarks and charged leptons are expressed as $v y_{u,1}^\dagger/\sqrt{2}$, $v y_{d,1}/\sqrt{2}$ and $v y_{e,1}/\sqrt{2}$, respectively. 
By the unitary transformations of fermions $\primed{f}_{L,R}=U_{L,R}f_{L,R}$, these matrices are diagonalized with real and positive eigenvalues. 
In the mass basis, the Yukawa interactions are rewritten as 
	%%%   yukawa term
		\begin{align}
			\mathcal{L}_\textrm{yukawa}=
				&	-\bar{Q}_L^u \rbra{
						\sqrt{2}\frac{M_u}{v}\tilde{\Phi}_1 +\rho_u\tilde{\Phi}_2
					} u_R
					-\bar{Q}_L^d \rbra{
						\sqrt{2}\frac{M_d}{v}\Phi_1 +\rho_d\Phi_2
					}d_R
				\nn\\
				&	-\bar{L}_L \rbra{
						\sqrt{2}\frac{M_e}{v}\Phi_1 +\rho_e\Phi_2
					} e_R
					+h.c.,
		\end{align}
where $Q_L^u=(u_L,V_\textrm{CKM}d_L)^T$ and $Q_L^d=(V_\textrm{CKM}^\dagger u_L,d_L)^T$ with $V_\textrm{CKM}$ being the Cabibbo-Kobayashi-Maskawa (CKM) matrix. 
The fermion-mass matrices $M_f$ are diagonal, while $\rho_f$ are $3\times3$ complex matrices whose off-diagonal elements generally induce dangerous FCNCs which are strongly constrained by flavor experiments.

In order to remove new contribution to the FCNCs at tree revel, we consider the alignment for the Yukawa sector proposed by Pich and Tuzon\cite{Pich:2009sp} as
	%%%   yukawa alignment
		\eq{
			y_{f,2}=\zeta_f ~y_{f,1}
		,}
where $\zeta_f$ are complex parameters. 
Thus, $y_{f,1}$ and $y_{f,2}$ are diagonalized at the same time, and then $\rho_f$ are expressed as
	%%%   extra yukawa matrix
		\texteq{
			\rho_u=\frac{\sqrt{2}}{v}M_u\zeta_u^*
		} and
		\texteq{	
			\rho_{d,e}=\frac{\sqrt{2}}{v}M_{d,e}\zeta_{d,e}
		.}
We mention another prescription to avoid the FCNCs suggested by Glashow and Weinberg\cite{Glashow:1976nt}, in which a $Z_2$ symmetry is imposed to the Higgs sector to forbid one of the Yukawa matrices for each type of fermions. 
In this case, four types of Yukawa interactions appear depending on the $Z_2$ charge assignment for the right-handed fermions\cite{Barger:1989fj,Aoki:2009ha}. 
The $\zeta_f$ factors of each model are summarized in Tab.~\ref{tb:zeta}. 
	%%%   tab: zeta
		\begin{table}
			\centering
			\begin{tabular}{|c||c|c|c|} \hline
				Model			& $\zeta_u$ &$\zeta_d$ & $\zeta_l$ \\ \hline \hline
				Our model		& arbitrary complex & arbitrary complex & arbitrary complex \\ \hline
				Type-I THDM		& $1/\tan\beta$ & $1/\tan\beta$ & $1/\tan\beta$ \\
				Type-II THDM		& $1/\tan\beta$ & $-\tan\beta$ & $-\tan\beta$ \\
				Type-X THDM	& $1/\tan\beta$ & $1/\tan\beta$ & $-\tan\beta$ \\
				Type-Y THDM		& $1/\tan\beta$ & $-\tan\beta$ & $1/\tan\beta$ \\ \hline
			\end{tabular}
			\caption{The $\zeta_f$ factors in the THDMs. The names of the models are referred to Refs.~\cite{Pich:2009sp,Aoki:2009ha}.}
			\label{tb:zeta}
		\end{table}

In the basis of the mass eigenstates of the fermions and the Higgs bosons, the Yukawa interaction terms are expressed as
	%%%   yukawa term in mass basis
		\begin{align}
			\mathcal{L}_\textrm{yukawa}^\textrm{int} =
					&	-\sum_{f=u,d,e}
						\cbra{
						\sum_{j=1}^3\bar{f}_L \rbra{\frac{M_f}{v} \kappa_f^j} f_R H^0_j+h.c. 
						}
					\nn\\
					&	-\frac{\sqrt{2}}{v}
						\cbra{
							-\zeta_u \bar{u}_R (M_u^\dagger V_\textrm{CKM}) d_L
							+\zeta_d \bar{u}_L (V_\textrm{CKM} M_d) d_R
							+\zeta_e \bar{\nu}_L M_e e_R
						} H^{+}+h.c. 
		,\end{align}
where $\kappa_f^j$ are the coupling factors for the interactions of the neutral Higgs bosons to the fermions which are given as
	%%%   kappa
		\begin{align}
			\kappa_f^j=\mathcal{R}_{1j}
						+\sbra{
							\mathcal{R}_{2j} +i(-2I_f)\mathcal{R}_{3j}
						} |\zeta_f|e^{i(-2I_f)\theta_f}
		,\end{align}
with $\theta_f\equiv\arg[\zeta_f]\in(-\pi,\pi]$ and $I_u=1/2, I_d=I_e=-1/2$. 
In our model, due to the alignment of the Higgs potential $\mathcal{R}_{jk}=\delta_{jk}$, the factors $\kappa_f^j$ are expressed as
	%%%   kappa
		\begin{align}
			\kappa_f^{1} &=1
		,\\	\kappa_f^{2} &=|\zeta_f| e^{i(-2I_f)\theta_f}
		,\\	\kappa_f^{3} &=i(-2I_f) \kappa_f^2
		.\end{align}
We can see that the Yukawa couplings for $H^0_1$ are real at tree level, while those of $H^0_{2,3}$ contain the CP-violating phases. 

%%%%%     kinetic term   %%%%%
\subsection{Kinetic Terms of the Scalar Fields}
The kinetic term of the scalar doublet fields is given as
	%%%   kinetic term
		\begin{align}
			\mathcal{L}_\textrm{kin}
				=|D_\mu\Phi_1|^2+|D_\mu\Phi_2|^2
		.\end{align}
The covariant derivative $D_\mu$ is given as
	%%%   covariant derivative for doublets
		\texteq{
			D_\mu=\partial_\mu+i g_2 \frac{\sigma^a}{2} W_\mu^a +i g_1 \frac{1}{2} B_\mu
		,}
where $g_2$ and $g_1$ are the SU(2)$_L$ and U(1)$_Y$ gauge coupling constants, respectively. 
In the mass eigenstates of the gauge bosons and Higgs bosons, the trilinear Higgs-gauge-gauge type couplings are given as
	%%%   HVV couplings
		\begin{align}
			\mathcal{L}_\textrm{kin}\supset
				\sum_j^3\mathcal{R}_{1j}
					\rbra{
						\frac{2m_W^2}{v}W_\mu W^\mu
						+\frac{m_Z^2}{v}Z_\mu Z^\mu
					} H^0_j
		,\end{align}
where the masses of the gauge bosons $m_V$ ($V=W,~Z$) are defined as $m_W=g_2v/2$ and $m_Z=\sqrt{g_2^2+g_1^2}~v/2$. 
When the alignment limit $\mathcal{R}_{jk}=\delta_{jk}$ is taken, the couplings of $H^0_1VV$ are the same as those of the SM, and the interactions of $H^0_2VV$ and $H^0_3VV$ vanish at tree level. 

%%%%%     theoretical and experimental constraints   %%%%%
\subsection{Theoretical and Experimental Constraints}
The dimensionless parameters of the Higgs potential are constrained by the perturbative unitarity\cite{Ginzburg:2005dt,Kanemura:1993hm,Akeroyd:2000wc,Kanemura:2015ska} and the vacuum stability\cite{Nie:1998yn,Kanemura:1999xf}. 
These constraints can be translated into those of the masses and the mixings. 
The masses and mixings of the Higgs bosons are also constrained by the electroweak $S$, $T$ and $U$ parameters\cite{Peskin:1990zt,Peskin:1991sw}. 
New contributions to the $T$ parameter from the additional Higgs bosons vanish at one-loop level by imposing $m_{H^0_3}=m_{H^\pm}$ and $\lam_6=0$, because of the custodial symmetry in the Higgs potential\cite{Pomarol:1993mu,Haber:1992py,Grzadkowski:2010dj,Haber:2010bw,Kanemura:2011sj}. 
In addition, the constraints from the $B$ physics experiments are taken into account\cite{Mahmoudi:2009zx}. 
In particular, we refer to Ref.~\cite{Jung:2010ab} for the constraint on $m_{H^\pm}$ and $\zeta_q$. 
For instance, for $m_{H^\pm}=200$ GeV, the upper limit of $|\zeta_u^*\zeta_d|$ is given to be $0.32 ~(2.4)$ for $\arg[\zeta_u^* \zeta_d]=0 ~(1)$. 
The EDM experiments constraining new CP-violating effects are discussed in next section. 
Finally, we comment on constraints from direct searches for additional Higgs bosons at the LHC, which provides upper limits on the cross section times branching ratio. 
We confirm that in our benchmark scenario which is defined in the next section, our prediction of the cross section times branching ratio is typically two orders of magnitude smaller than the upper limit\cite{Aad:2020zxo}.

%%%%%%%%%%     EDM     %%%%%%%%%%
\section{Scenario cancelling the electric dipole moment}\label{sc:EDM}
We review the scenario without large EDMs which has been studied in the previous work\cite{Kanemura:2020ibp}.

%%%%%     constraint     %%%%%
\subsection{Constraint from the Electric Dipole Moment Data}
The EDM $d_f$ of a fermion $f$ is defined by the effective Lagrangian as
	%%%   EDM Lagrangian
		\begin{align}
			\mathcal{L}_\textrm{EDM}=-\frac{d_f}{2}\bar{f} \sigma^{\mu\nu}(i\gamma^5)f F_{\mu\nu}
		,\end{align}
where $F_{\mu\nu}$ is the electromagnetic field strength tensor and
	%%% sigma mu nu
		\texteq{
			\sigma^{\mu\nu}=\frac{i}{2}[\gamma^\mu,\gamma^\nu]
		.}
For the neutron EDM, there are also the contribution from the chromo-EDM (CEDM) of a quark defined as
	%%%   CEDM Lagrangian
		\begin{align}
			\mathcal{L}_\textrm{CEDM}=-\frac{d_q^C}{2}\bar{q} \sigma^{\mu\nu}(i\gamma^5)q G_{\mu\nu}
		,\end{align}
where the $G_{\mu\nu}$ is the QCD field strength tensor. 

The most severe constraint on $d_e$ has been given by the ACME collaboration using the thorium-monoxide EDM which gives the upper limit $|d_e+kC_S|<1.1\times10^{-29}$ e~cm at the 90\% confidence level (CL)\cite{Andreev:2018ayy}. 
The second term $C_S$ is defined as the coefficient of the dimension six operator for the electron-nucleon interaction given as
	%%%   (e gamma5 e)(NN)
		\texteq{
			\mathcal{L}\supset C_S(\bar{e}i\gamma_5 e)(\bar{N}N)
		.}
The coefficient $k$ is given as $k\sim \mathcal{O}(10^{-15})$ GeV${}^2$ e~cm\cite{Cheung:2014oaa}. 
According to the discussion given in Refs.~\cite{Cheung:2014oaa,Jung:2013hka}, the value of $kC_S$ is typically two orders of magnitude smaller than the current bound in our benchmark scenario which is introduced below. 
Therefore, we neglect this contribution, and we simply consider the bound on the electron EDM as $|d_e|<1.1\times10^{-29}$ e~cm (90\% CL) in the following discussion. 
In our previous paper\cite{Kanemura:2020ibp}, we confirmed that the neutron EDM $d_n$ does not give stringent constraints, so that we focus on the constraint from $d_e$ in the following discussion.

%%%%%     BZ contributions     %%%%%
\subsection{Destructive Interference between the Barr-Zee Type Contributions}
The dominant contributions to $d_f$ are given by the two-loop BZ type diagrams\cite{Barr:1990vd}, while the one-loop contributions are negligibly smaller than them because the one-loop diagram contains two additional powers of small Yukawa couplings for light fermions compared with the BZ diagrams. 
From the discussion in Ref.~\cite{Bernreuther:1990jx}, we can estimate $d_e(\textrm{1-loop})\sim \mathcal{O}(10^{-34})$ e~cm, while $d_e(\textrm{2-loop})\sim \mathcal{O}(10^{-27})$ e~cm in the typical input parameters with $\mathcal{O}(100)$ GeV of the masses of extra Higgs bosons and the $\kappa_f$ factors of $\mathcal{O}(1)$.\footnote{
In addition to the differences of the power of small Yukawa couplings and the loop factor between one loop and BZ contributions, a difference of loop functions and the additional factor $e^2$ given in the BZ contribution provide a factor of $\mathcal{O}(100)$ in the ratio of the one loop to the BZ contribution\cite{Bernreuther:1990jx}. 
Thus, one loop contributions are typically $\mathcal{O}(10^{-7})$ smaller than two loop BZ contributions. 
} 
Therefore, the BZ type contribution is typically two orders of magnitude larger than the current upper limit of $|d_e|$, and some cancellations are required to satisfy the data. 

\begin{figure}
			\begin{tabular}{c}
				%%%
				\begin{minipage}{0.5\hsize}
					\centering
					\includegraphics[width=50mm]{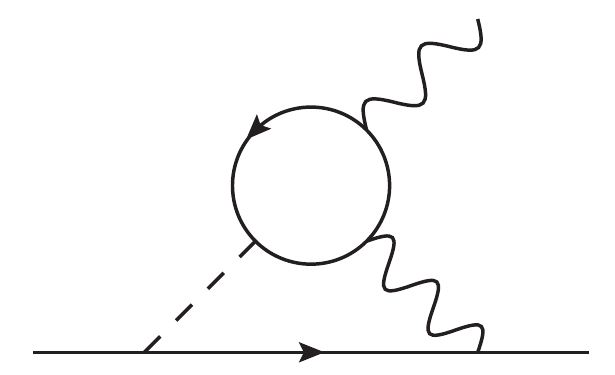}
					\\(\ref{fg:BZ}-1) Fermion-loop
				\end{minipage}
				%%%
				\begin{minipage}{0.5\hsize}
					\centering
					\includegraphics[width=50mm]{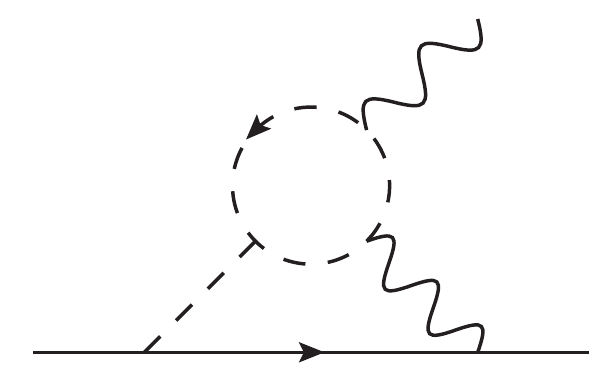}
					\\(\ref{fg:BZ}-2) Higgs boson-loop
				\end{minipage}
				%%%
			\end{tabular}
			\caption{BZ type diagrams giving the dominant contribution to the electron EDM under the alignment for the Higgs potential in which the contribution from the gauge boson loop vanishes.}
			\label{fg:BZ}
		\end{figure}

In the alignment scenario, there are two types of contributions to the BZ diagram shown in Fig.~\ref{fg:BZ}. 
Thus, the BZ contribution to $d_f$ can be expressed by the type of particles in the loop as
	%%%   BZ EDM
		\eq{
			d_f=d_f(\textrm{fermion})+d_f(\textrm{Higgs})%+d_f(\textrm{gauge})
		\label{eq:BZcontribution1}
		.}
We note that gauge boson-loops $d_e(\textrm{gauge})$ do not contribute to the BZ diagram, because they are proportional to $\sum_{j=1}^3\mathcal{R}_{1j}\im[\kappa_f^j]$ which becomes zero in the alignment limit $\mathcal{R}_{ij}=\delta_{ij}$. 
Each contribution can be further classified by the intermediated gauge boson as
		\eq{
			d_f(X)=d_f^\gamma(X)+d_f^Z(X)+d_f^W(X)
		\label{eq:BZcontribution2}
		.}
The explicit expression of each contribution $d_f^V(X)$ ($V=\gamma, Z$ and $W$) in the THDM with the alignment of the Yukawa interaction are given in Appendix~\ref{sc:BZformulae}. 
We note that non-BZ type diagrams at two-loop level mediated by Higgs bosons and gauge bosons are also proportional to $\sum_{j=1}^3\mathcal{R}_{1j}\im[\kappa_f^j]$\cite{Leigh:1990kf,Jung:2013hka,Altmannshofer:2020shb}, so that they vanish in the alignment limit. 
	%%%   fig: BZ diagram

In Ref.~\cite{Kanemura:2020ibp}, we have considered the destructive interference between the contributions from the fermion-loop and Higgs boson-loop. 
Requiring $d_e\simeq0$, a relation for the extra coupling constants and the CP-violating phases appears. 
When the extra Higgs boson masses are nearly degenerate and $\theta_d=0$ is assumed, such a relation is given as
	%%%   destructive EDM
		\begin{align}
			\rbra{A^\gamma I^\gamma+A^Z I^Z+A^W I^W}|\zeta_u| \sin(\theta_u-\theta_e)
			\simeq
			-\rbra{B^\gamma J^\gamma+B^Z J^Z+B^W J^W}|\lam_7| \sin(\theta_7-\theta_e)
		\label{eq:destructiveEDM}
		,\end{align}
where $A$ and $B$ are the constant factors:
	%%%
		\begin{align}
				A^\gamma	&=\frac{32}{3}\sin^2\theta_W\frac{m_t^2}{v^2}
			,\\	A^Z			&=\frac{1}{3}\frac{(1-4\sin^2\theta_W) (3-8\sin^2\theta_W)}{\cos^2\theta_W}\frac{m_t^2}{v^2}
			,\\	A^W		&=3\frac{m_t^2}{v^2}
			,\\	B^\gamma	&=4\sin^2\theta_W
			,\\	B^Z			&=\frac{1}{2}\frac{(1-4\sin^2\theta_W)}{\cos^2\theta_W} \cos{2\theta_W}
			,\\	B^W			&=-\frac{1}{2}
		,\end{align}
and $I$ and $J$ are the loop functions depending on the masses of the extra Higgs bosons:
	%%%
		\begin{align}
				I^V	&=2\int_0^1dz \sbra{\frac{1}{z}-(1-z)} C^{V\tilde{H}}_{tt}(z)
				,\quad\quad(V=\gamma~\textrm{and}~Z)
			,\\	I^W	&=\int_0^1dz \frac{2-z}{z}\sbra{\frac{2}{3}-z} C^{W\tilde{H}}_{tb}(z)
			,\\	J^V	&=2\int_0^1dz (1-z) C^{V\tilde{H}}_{\tilde{H}\tilde{H}}(z)
				,\quad\quad\quad(V=\gamma,Z~\textrm{and}~W)
		,\end{align}
where $C^{GH}_{XY}(z)$ are given in Appendix~\ref{sc:BZformulae}, and its argument $m_{\tilde{H}}$ corresponds to the typical mass of the additional Higgs bosons. 
From Eq.~\eqref{eq:destructiveEDM}, it is seen that the independent phases $\theta_u$, $\theta_e$ and $\theta_7$ can be taken such that the fermion- and the Higgs boson-loop contributions to $d_e$ cancel with each other. 

In Ref.~\cite{Kanemura:2020ibp}, we found the benchmark parameter point which satisfies the electron EDM data by using the SM input parameters shown in Tab.~\ref{tb:SMinput}. 
The new input parameters of the THDM are also shown in Tab.~\ref{tb:THDMinput}. 
	%%%   tab: inputs of SM para
		\begin{table}
			\centering
			\begin{tabular}{|cccc|}
				\hline
					$m_u=1.29\times10^{-3}$,
				&	$m_c=0.619$,
				&	$m_t=171.7$,
				&
				\\
					$m_d=2.93\times10^{-3}$,
				&	$m_s=0.055$,
				&	$m_b=2.89$,
				&
				\\
					$m_e=0.487\times10^{-3}$,
				&	$m_\mu=0.103$,
				&	$m_\tau=1.746$
				&	~~(in GeV)
				\\ \hline
			%\end{tabular}
			%\\
			%\begin{tabular}{cccc}
					$\alpha_\textrm{em}=1/127.955$,
				&	$m_Z=91.1876$ GeV,
				&	$m_W=80.379$ GeV,
				&	$\alpha_S=0.1179$
				\\ \hline
			%\end{tabular}
			%\\
			%\begin{tabular}{cccc}
					$\lambda=0.22453$,
				&	$A=0.836$,
				&	$\bar{\rho}=0.122$,
				&	$\bar{\eta}=0.355$
				\\ \hline
			\end{tabular}
			\caption{
				Input values for the SM parameters at the $m_Z$ scale. 
				The fermion masses and the other parameters are taken from Refs.~\cite{Xing:2007fb,Bijnens:2011gd} and Ref.~\cite{Tanabashi:2018oca}, respectively. 
				The Wolfenstein parameters of the CKM matrix $\lambda$, $A$, $\bar{\rho}$ and $\bar{\eta}$ are defined in Ref.~\cite{Tanabashi:2018oca}. 
			}
			\label{tb:SMinput}
		\end{table}
In this benchmark point, we obtain $|d_e|=9.5\times10^{-30}$ e~cm which is just below the current experimental limit. 
As we discussed in Ref.~\cite{Kanemura:2020ibp}, the value of $|d_e|$ is stable until about $10^7$ GeV by using the renormalization equations at one loop level. 
In addition, we confirmed that the Landau pole does not appear until about $10^{10}$ GeV. 
	%%%   tab: inputs of THDM para
		\begin{table}
			\centering
			\begin{tabular}{|ccccc|}
				\hline
					$M=240$,
				&	$m_{H^0_2}=280$,
				&	$m_{H^0_3}=230$,
				&	$m_{H^\pm}=230$
				&	(in GeV)
				\\ \hline
					$|\zeta_u|=0.01$,
				&	$|\zeta_d|=0.1$,
				&	$|\zeta_e|=0.5$,
				&	$|\lam_7|=0.3$,
				&	$\lam_2=0.5$
				\\ \hline
					$\theta_u=1.2$,
				&	$\theta_d=0$,
				&	$\theta_e=\pi/4$,
				&	$\theta_7=-1.8$
				&	(in radian)
				\\ \hline
			\end{tabular}
			\caption{
				Input values of the benchmark point taken from the previous paper\cite{Kanemura:2020ibp} for the THDM at the scale $m_Z$. 
				}
			\label{tb:THDMinput}
		\end{table}

In the following, we discuss the allowed parameter region by $|d_e|$ around the benchmark point. 
We scan the $\theta_7$ parameter with $(-\pi,\pi]$ and the $|\lam_7|$ parameter to be larger than $0.01$, because they do not affect the discussion of the collider phenomenology given in the next section.

\begin{figure}
		\centering
			\begin{tabular}{c}
				\begin{minipage}{0.33\hsize}
					\centering
					\includegraphics[height=55 mm]{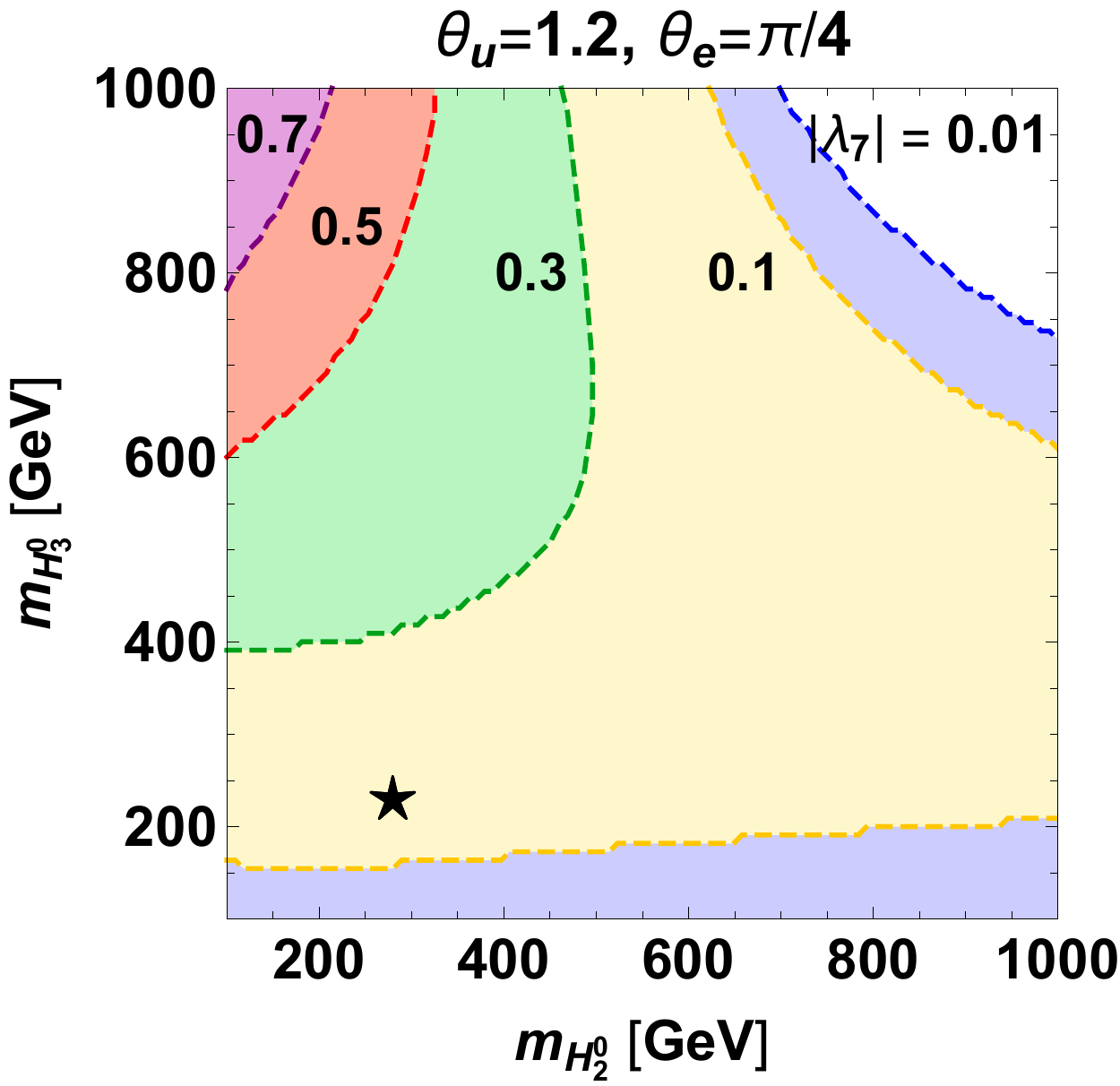}
					%\\(\ref{fg:}-a)
				\end{minipage}
				%%%
				\begin{minipage}{0.33\hsize}
					\centering
					\includegraphics[height=55 mm]{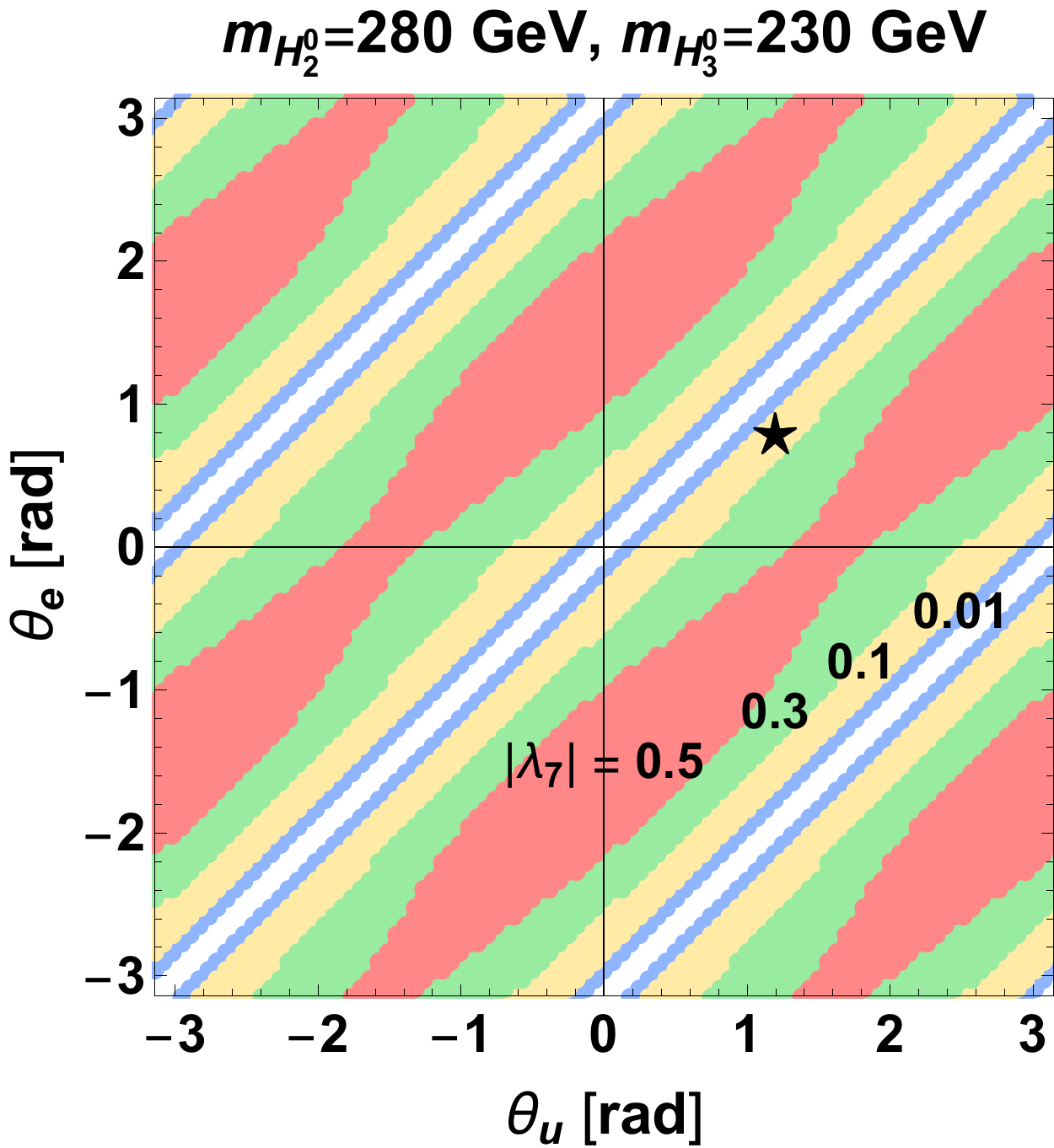}
					%\\(\ref{fg:}-b)
				\end{minipage}
				%%%
				\begin{minipage}{0.33\hsize}
					\centering
					\includegraphics[height=55 mm]{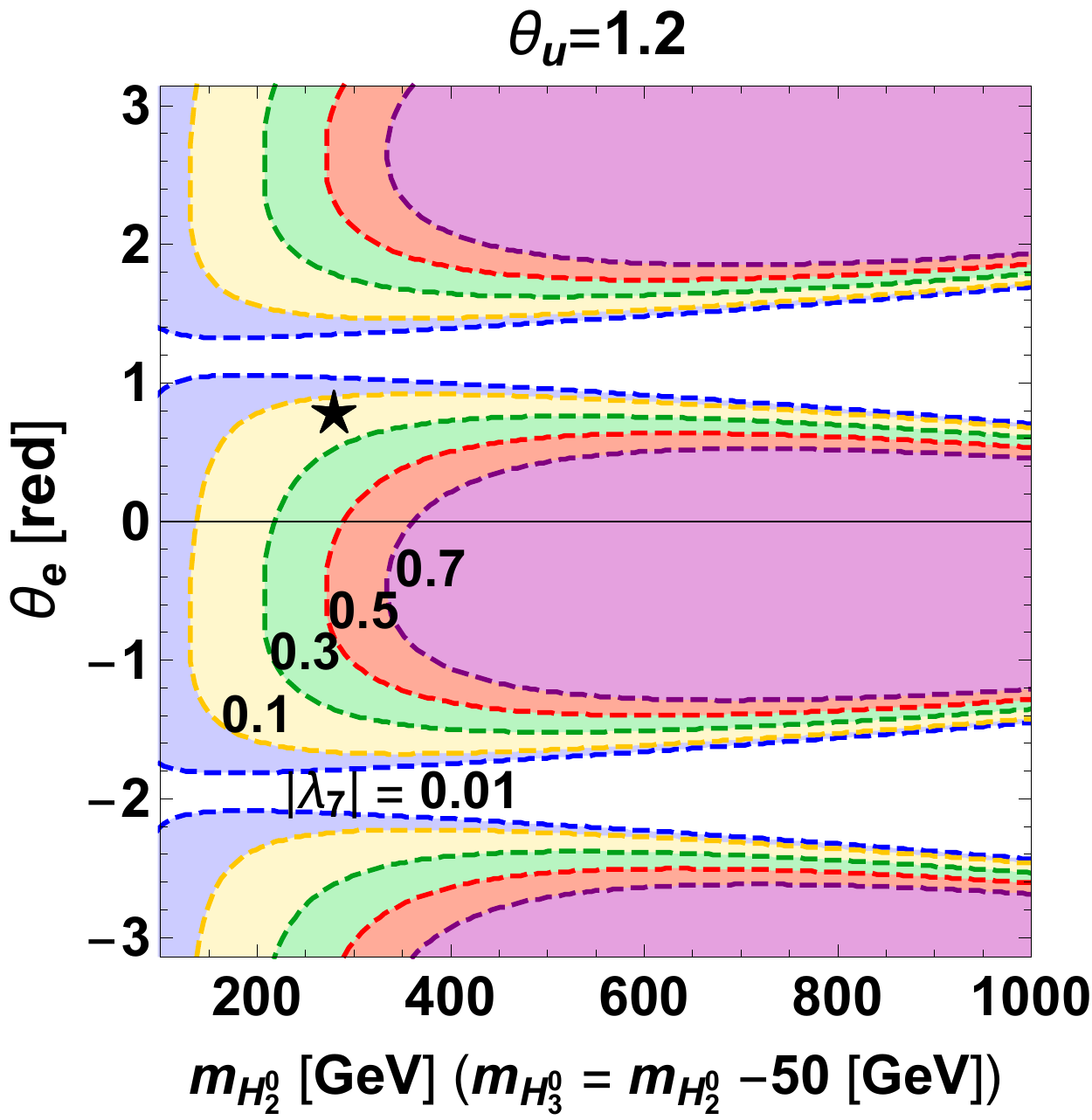}
					%\\(\ref{}-c)
				\end{minipage}
				%%%
			\end{tabular}
		\caption{
			Excluded region against for the electron EDM data on the $m_{H^0_2}$-$m_{H^0_3}$ (left), $\theta_u$-$\theta_e$ (center) and $m_{H^0_2}$-$\theta_e$ (right) plane under the scan of $\theta_7\in(-\pi,\pi]$. 
			We take $\theta_u=1.2$ and $\theta_e=\pi/4$ in the left panel, $m_{H^0_2}=280$ GeV and $m_{H^0_3}=230$ GeV in the center panel, $\theta_u=1.2$ and $m_{H^0_3}=m_{H^0_2}-50$ GeV in the right panel.
			In all the panels, we fix $|\zeta_u|=0.01$, $|\zeta_d|=0.1$, $|\zeta_e|=0.5$ and $\theta_d=0$. 
			The blue, yellow, green, red and purple shaded regions are excluded when $|\lam_7|$ is fixed to be $0.01$, $0.1$, $0.3$, $0.5$ and $0.7$, respectively, where the colored region with a smaller value of $|\lam_7|$ includes that with larger $|\lam_7|$. 
			The point marked by the star corresponds to the benchmark point when we choose $\theta_7=-1.8$ and $|\lambda_7|=0.3$. 
			}
		\label{masstheta}
	\end{figure}

In Fig.~\ref{masstheta}, we show the excluded region against for the electron EDM data on the $m_{H^0_2}$-$m_{H^0_3}$ (left), $\theta_u$-$\theta_e$ (center) and $m_{H^0_2}$-$\theta_e$ (right) plane under the scan of $\theta_7\in(-\pi,\pi]$. 
Each colored region is excluded for a fixed value of $|\lam_7|$ to be $0.01$ (blue), $0.1$ (yellow), $0.3$ (green), $0.5$ (red) and $0.7$ (purple). 
We note that in the center panel all the regions are allowed when $|\lam_7|\geq0.7$. 
The benchmark point is indicated by the star when we choose $\theta_7=-1.8$ and $|\lam_7|=0.3$. 
We can see that our benchmark point is allowed for $|\lam_7|\gtrsim0.2$ with an appropriate value of $\theta_7$. 
From the three panels, we can see that larger parameter regions are excluded for smaller $|\lam_7|$, in which the Higgs boson-loop contribution cannot be large enough to cancel the fermion-loop contribution. 
The behavior of each panel can be understood as follows. 
In the left panel, the region with larger $m_{H^0_2}$ and $m_{H_3^0}$ is allowed for smaller values of $|\lam_7|$, which is simply because of the decoupling of the new contribution to $|d_e|$. 
For smaller values of $m_{H_2^0}$, we see that a larger value of $|\lam_7|$ is required for larger $m_{H^0_3}$. 
This can be understood by the fact that the Higgs boson-loop contribution becomes smaller when $m_{H^\pm}(=m_{H^0_3})$ is larger, while the fermion-loop contribution does not become smaller so much. 
Thus, a larger value of $|\lam_7|$ is needed to compensate for the reduction of the Higgs boson-loop effect. 
In the center panel, the region with $\theta_e \simeq \theta_u \pm \pi n$ ($n = 0,1,2,\dots$) is allowed even for smaller $|\lam_7|$, because the fermion-loop contribution is proportional to $\sin(\theta_u - \theta_e)$, see Eq.~\eqref{eq:destructiveEDM}. 
In the region apart from the above case, the fermion-loop becomes significant, and the cancellation from the Higgs boson-loop is necessary with appropriate value of $|\lam_7|$. 
The behavior of the right panel can be understood in a similar way to the center panel. 
Namely, the region with satisfying $\theta_e \simeq \theta_u \pm \pi n$, i.e., $\theta_e \simeq 1.2$ and $-1.9$ for $\theta_u = 1.2$ is allowed even for smaller $|\lam_7|$. 
If we consider the region apart from the above case, a larger value of $|\lambda_7|$ is required to satisfy the constraint from the electronEDM data. 
For a fixed value of $\theta_e$, we see that the required value of $\lam_7$ becomes monotonically larger when $m_{H^0_2}$ is larger due to the fixed relation of $m_{H^0_3}(=m_{H^\pm}) = m_{H^0_2} - 50$ GeV. 
In addition, we also see that the allowed regions with the smaller values of $|\lam_7|$ become wider for the larger $m_{H^0_2}$ due to the decoupling of the new contribution to $|d_e|$. 

%%%%%%%%%%     measurement of CP violation     %%%%%%%%%%
\section{Measurement of CP property}\label{sc:CPV}
\subsection{Decay of the Extra Higgs Bosons}\label{decay}
We discuss the branching ratios of the extra Higgs bosons. 
Since we assume the alignment limit, i.e., $\mathcal{R}_{jk}=\delta_{jk}$, the extra Higgs bosons mainly decay into a fermion pair. 
If it is kinematically allowed, they can also decay into a (off-shell) gauge boson and another Higgs boson. 
In addition, there are loop-induced decay processes such as $H^0_{2,3}\to \gamma\gamma$, $Z\gamma$, $gg$ and $H^\pm\to W^\pm Z$, $W^\pm\gamma$. 
Except for the $H^0_{2,3}\to gg$, these branching ratios are negligibly small; i.e., $BR(H^0_{2,3}\to \gamma\gamma/Z\gamma)$ and $BR(H^\pm\to W^\pm Z, W^\pm\gamma)$\cite{CapdequiPeyranere:1990qk,Kanemura:1997ej} are typically smaller than $\mathcal{O}(10^{-4})$. 
The analytic formulae of the decay rates are given in Appendix~\ref{sc:decayrate}. 

In Fig.~\ref{fg:BR}, we show the branching ratios of $H^0_2$ (left), $H^0_3$ (center) and $H^\pm$ (right) as a function of the extra-Higgs-boson mass with $m_{H^0_2}-50$ GeV $=m_{H^0_3}=m_{H^\pm}$ and the other parameters being the same as the benchmark point given in Tab.~\ref{tb:THDMinput}. 
For $H^0_3$, the dominant decay modes are $\tau^+\tau^-$/$b\bar{b}$ ($t\bar{t}$) with their branching ratios to be about 50 (80) \% when the mass of $H^0_3$ is smaller (larger) than $2m_t$. 
On the other hand, $H^0_2$ mainly decays into $Z^*H^0_3$ and $W^\pm{}^*H^\mp$ due to the mass difference, while the branching ratios of $H^0_2\to\tau^+\tau^-$ and $b\bar{b}$ can be $\mathcal{O}(10)$\% for $m_{H^0_2}<2m_t$. 
For $H^\pm$, the main decay mode is $\tau\nu$ ($tb$) if $m_{H^\pm}$ is smaller (larger) than the top mass. 
	%%%   fig: branching ratio
		\begin{figure}
					\centering
					\includegraphics[width=53 mm]{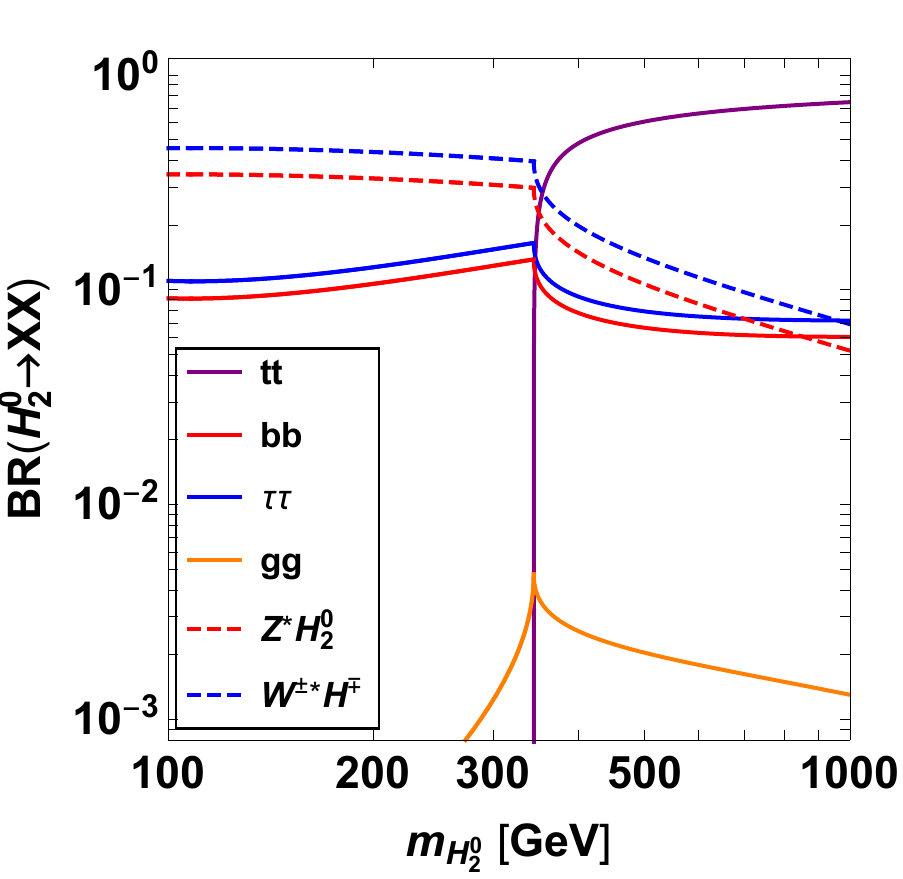}
					\includegraphics[width=53 mm]{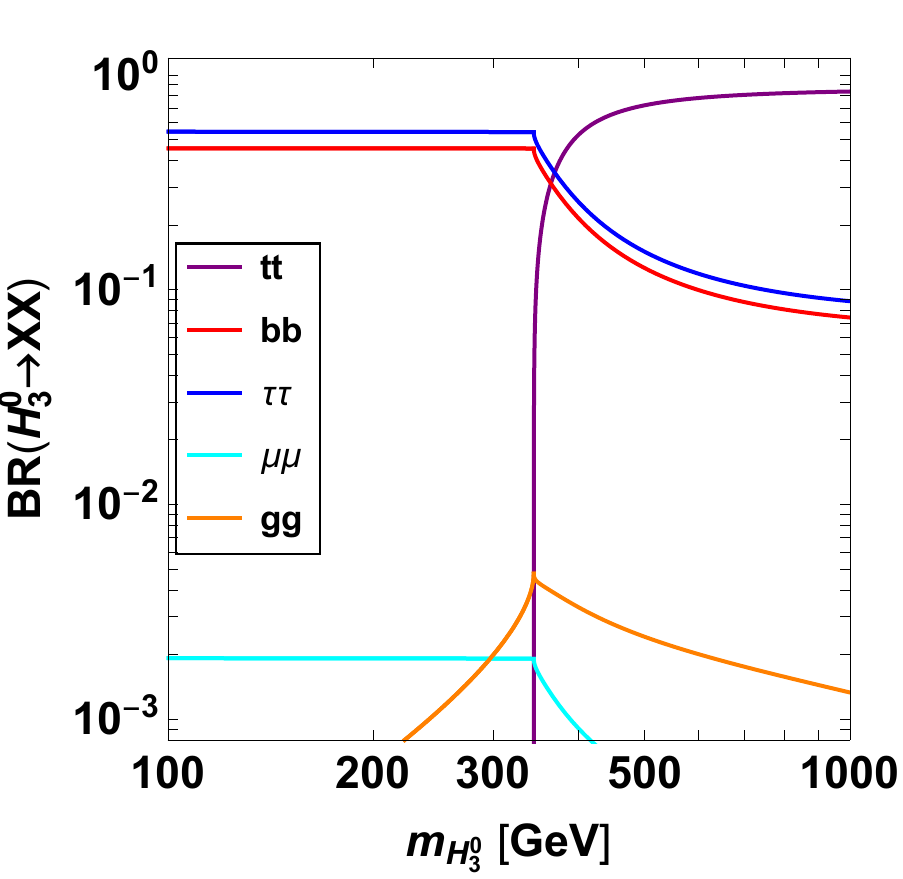}
					\includegraphics[width=53 mm]{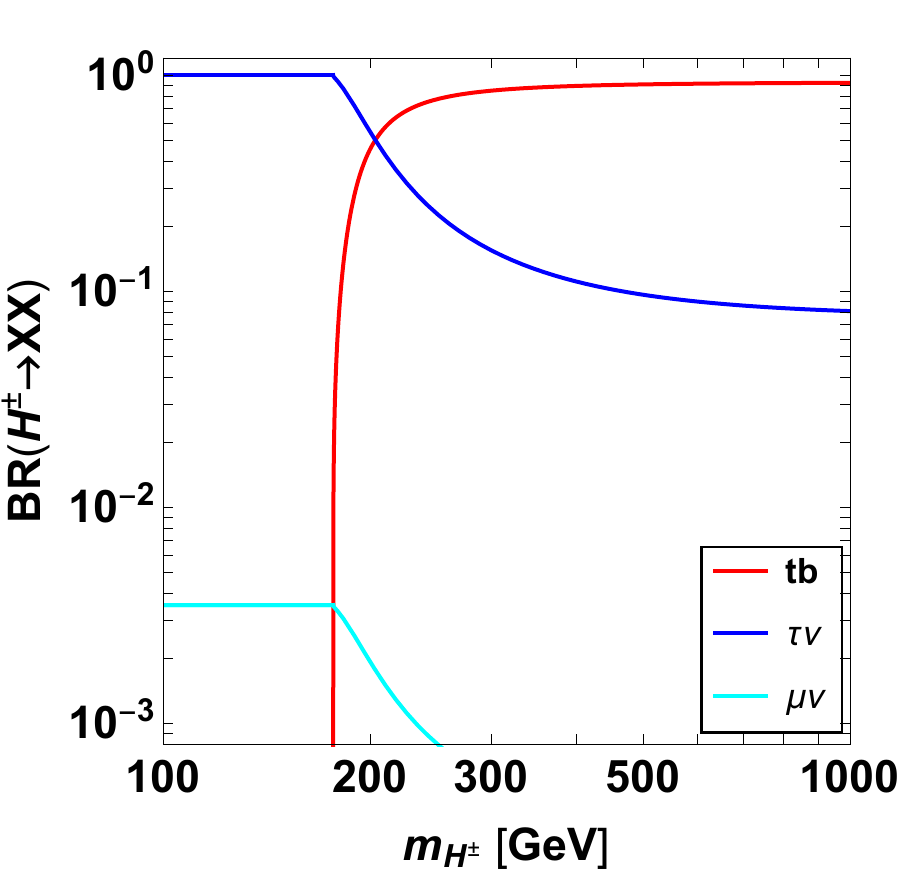}
			\caption{	
				Branching ratios of $H^0_2$ (left), $H^0_3$ (center) and $H^\pm$ (right) as a function of the extra-Higgs-boson mass with $m_{H^0_2}-50$ GeV $=m_{H^0_3}=m_{H^\pm}$ and the other parameters being the same as the benchmark point given in Tab.~\ref{tb:THDMinput}. 
			}
			\label{fg:BR}
		\end{figure}

\subsection{Angular Distribution}
In our scenario, effects of CP-violating phases of the Yukawa interaction terms appear in angular distributions of the decay products of the extra Higgs bosons. 
On the other hand, the CP-violating phases of the Higgs potential do not directly affect the decays of the extra Higgs bosons into fermions. 
However, by taking into account the necessity of the EDM cancellation as discussed in Sec.~\ref{sc:EDM}, 
the existence of the CP-violating phases in the Higgs potential can be indirectly proved by measuring the CP-violating effect in the angular distributions. 

We discuss the decay of the additional neutral Higgs bosons into a tau lepton pair which has relatively clearer signatures than the others. 
Hadronic decays of the tau lepton can be useful to extract the information of the CP-violating phase due to their simple kinematic structure\cite{Kuhn:1982di,Grzadkowski:1995rx,Hagiwara:2012vz,Harnik:2013aja,Jeans:2018anq}. 
We thus consider $H^0_j\to\tau^{-}\tau^{+}\to X^{-}\nu X^{+}\bar{\nu}$, where $X^\pm$ are hadrons, for instance, $\pi^\pm$, $\rho^\pm$ or $a_1^\pm$ mesons. 
The $\rho^\pm$ ($a_1^\pm$) mesons further decay into $\pi^\pm\pi^0$ ($\pi^\pm\pi^0\pi^0$ or $\pi^\pm\pi^\pm\pi^\mp$). 

\begin{figure}
			\centering
			\includegraphics[height=60 mm]{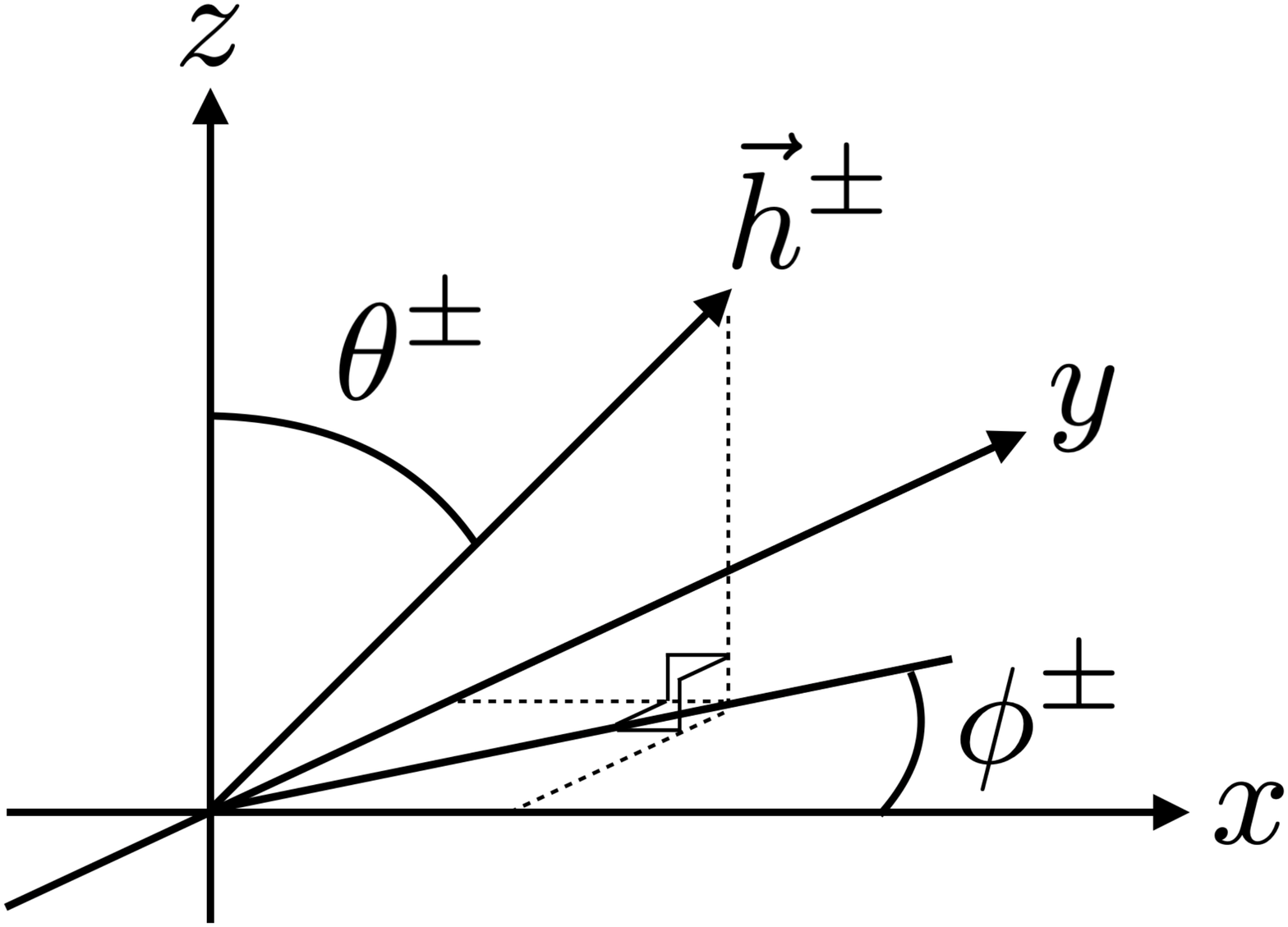}
			\caption{
				Schematic pictures for the angles $\theta^\pm$ and $\phi^\pm$ defined in the rest frame of $\tau^\pm$. 
				The $z$ axis is defined along with the direction of $\tau^\pm$ in the rest frame of $H^0_j$. 
				}
			\label{fg:angleDef}
		\end{figure}

\begin{figure}
			\begin{tabular}{c}
				\begin{minipage}{0.5\hsize}
					\centering
					\includegraphics[width=75 mm]{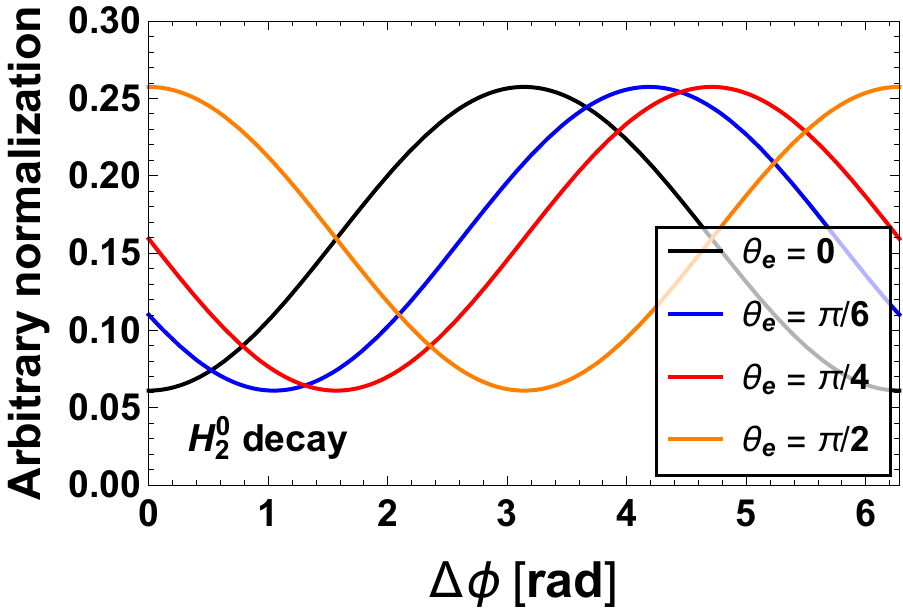}
					%\\(\ref{fg:angldistplot}-1)
				\end{minipage}
				%%%
				\begin{minipage}{0.5\hsize}
					\centering
					\includegraphics[width=75 mm]{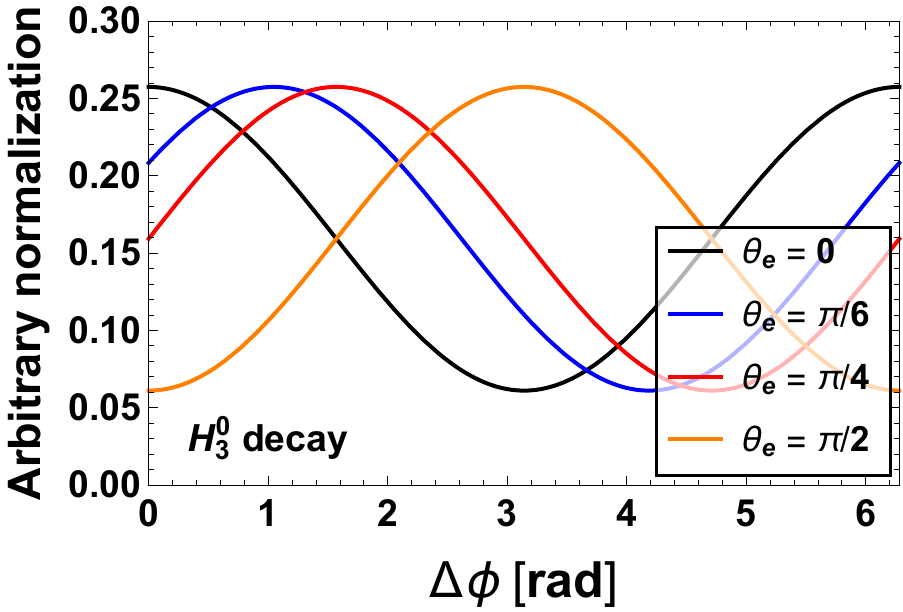}
					%\\(\ref{fg:angldistplot}-2)
				\end{minipage}
				%%%
			\end{tabular}
			\caption{
				Normalized $\Delta\phi$ distributions for $H^0_j\to\tau^{-}\tau^{+}\to X^{-}\nu X^{+}\bar{\nu}$ with $j=$2 (left) and 3 (right). 
				}
			\label{fg:angldistplot}
		\end{figure}

The squared amplitude for $H^0_j\to\tau^{-}\tau^{+}\to X^{-}\nu X^{+}\bar{\nu}$ is calculated as
	%%%
		\begin{align}
			\overline{|\mathcal{M}(\theta^-, \phi^-, \theta^+, \phi^+)|^{2}}\propto(1+\cos\theta^-\cos\theta^+)-\sin\theta^-\sin\theta^+\cos(2\theta_e-\Delta\phi)
		\label{eq:amplitude}
		,\end{align}
where the mass of the tau lepton is neglected and $\Delta\phi\equiv\phi^+-\phi^-$. 
The angles $\theta^\pm$ and $\phi^\pm$ are defined in association with the polarimeter $h^\pm_\mu$ in the rest frame of $\tau^\pm$ as depicted in Fig.\ref{fg:angleDef}. 
	%%%   fig: angle
The polarimeter $h^\pm_\mu$ is given by the momenta of the decay products of $\tau^\pm$ in the rest frame of $\tau^\pm$ as
	%%%
		\begin{align}
			\vec{h}^\pm \propto \vec{\Pi}^\pm-\vec{\Pi}^5{}^\pm
		,\end{align}
where
	%%%
		\begin{align}
				\Pi^\pm_\mu &\equiv 4\re(J^\pm_\mu k^\pm\cdot J^\pm)-2k^\pm_\mu J^\pm\cdot J^\pm{}^*
			,\\	\Pi^5_\mu{}^\pm &\equiv 2\epsilon_{\mu\nu\rho\sigma} \im(J^\pm{}^\nu {J^\pm{}^*}^\rho k^\pm{}^\sigma)
		,\end{align}
with $k_\mu^-$ ($k_\mu^+$) is the four momentum of the neutrino (anti-neutrino).
In the above expressions, $J^\pm_\mu$ is the hadronic current given as 
		\eq{
			&J^\pm{}^\mu \propto q_{\pi^\pm}^\mu,	\quad (\textrm{for }X=\pi^\pm)
		,\\	&J^\pm{}^\mu \propto q_{\pi^\pm}^\mu-q_{\pi^0}^\mu,	\quad (\textrm{for }X=\rho^\pm)
		,\\	&J^\pm{}^\mu \propto
				\sbra{q_1^\mu-q_3^\mu-Q^\mu\frac{Q\cdot(q_1-q_3)}{Q^2}} F\rbra{(q_1+q_3)^2}
				+(1\leftrightarrow 2),
				\quad (\textrm{for }X=a_1^\pm)
		,}
where $Q=q_1+q_2+q_3$ with ${q_1}^\mu$, ${q_2}^\mu$ and $q_3^\mu$ being the four momentum of $\pi^0$, $\pi^0$ and $\pi^\pm$ ($\pi^\pm$, $\pi^\pm$ and $\pi^\mp$), in the decay of $a_1^\pm\to\pi^0\pi^0\pi^\pm$ ($a_1^\pm\to\pi^\pm\pi^\pm\pi^\mp$).
The function $F(Q^2)$ is given by
	%%%
		\begin{align}
			F(Q^2)=\frac{B_\rho(Q^2)+\alpha B_{\rho'}(Q^2)}{1+\alpha}
		,\end{align}
with $\alpha=0.145$\cite{Hagiwara:2012vz} and the Breit-Wigner factor
	%%%
		\begin{align}
			B_V(Q^2)=\frac{m_V^2}{m_V^2-Q^2-i\sqrt{Q^2}\Gamma_V(Q^2)}
		.\end{align}
The running width is
	%%%
		\begin{align}
			\Gamma_V(Q^2)=\Gamma_V\frac{\sqrt{Q^2}}{m_V}
					\frac
						{\bar{\beta}^3\left(m_{\pi^-}^2/Q^2,m_{\pi^0}^2/Q^2\right)}
						{\bar{\beta}^3\left(m_{\pi^-}^2/m_V^2,m_{\pi^0}^2/m_V^2\right)}
		,\end{align}
where
	%%%
		\begin{align}
			\bar{\beta}(a,b)=(1+a^2+b^2-2a-2b-2ab)^{1/2}
		.\end{align}
By integrating out $\theta^\pm$ in Eq.~\eqref{eq:amplitude}, the angular distribution of the decay products of $H^0_{2,3}$ is obtained as follows
	%%%
		\begin{align}
			&\int\int d\cos\theta^- d\cos\theta^+ ~\overline{|\mathcal{M}(H^0_2\to\tau^{-}\tau^{+}\to X^{-}\nu X^{+}\bar{\nu})|^{2}}
								\propto16-\pi^2\cos(2\theta_e-\Delta\phi)
		,	\label{eq:angldist2}
		\\	&\int\int d\cos\theta^- d\cos\theta^+ ~\overline{|\mathcal{M}(H^0_3\to\tau^{-}\tau^{+}\to X^{-}\nu X^{+}\bar{\nu})|^{2}}
								\propto16-\pi^2\cos(2\theta_e-\Delta\phi+\pi)
			\label{eq:angldist3}
		,\end{align}
where the normalized $\Delta\phi$ distribution for the decay of $H^0_2$ ($H^0_3$) is plotted by the left (right) panel in Fig.~\ref{fg:angldistplot}. 
It is clear that the shape of the $\Delta\phi$ distribution strongly depends on the value of $\theta_e$. 
In the next section, we perform the signal and background simulation whether we can see the difference of the $\Delta\phi$ distribution at future lepton colliders. 
	%%%   fig: angldistplot

%%%%%%%%%%     collider signals     %%%%%%%%%%
\section{Simulation studies at the ILC}\label{sc:collider}
In this section, we discuss the testability of the scenario cancelling the EDM at future lepton colliders. 
In particular, we investigate the feasibility of measuring the effect of the CP-violating phases at the ILC. 
First, we show the production cross section of the extra neutral Higgs bosons, and define the signal process which contains information of the CP-violating phases. 
Second, we discuss corresponding background processes. 
Third, we demonstrate various distributions which are reconstructed from kinematic observables of the final state particles, and consider background reductions by imposing appropriate kinematical cuts. 
Finally, we give a brief comment on the testability of measuring the CP property at the LHC. 
Throughout the analysis, we assume that the additional Higgs bosons are discovered at future LHC experiments, and information of their masses are obtained to some extent.

\subsection{Production of the Extra Higgs Bosons}
\begin{figure}
		\centering
		\includegraphics[width=65 mm]{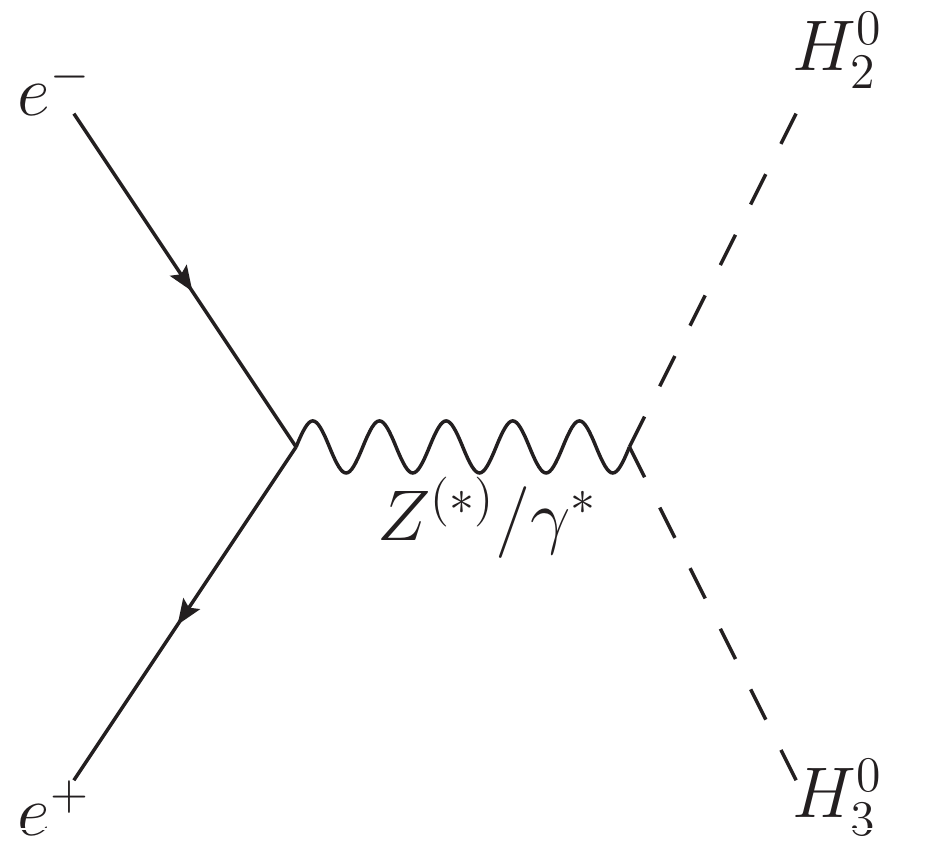}
		\hspace{10 mm}
		\includegraphics[width=65 mm]{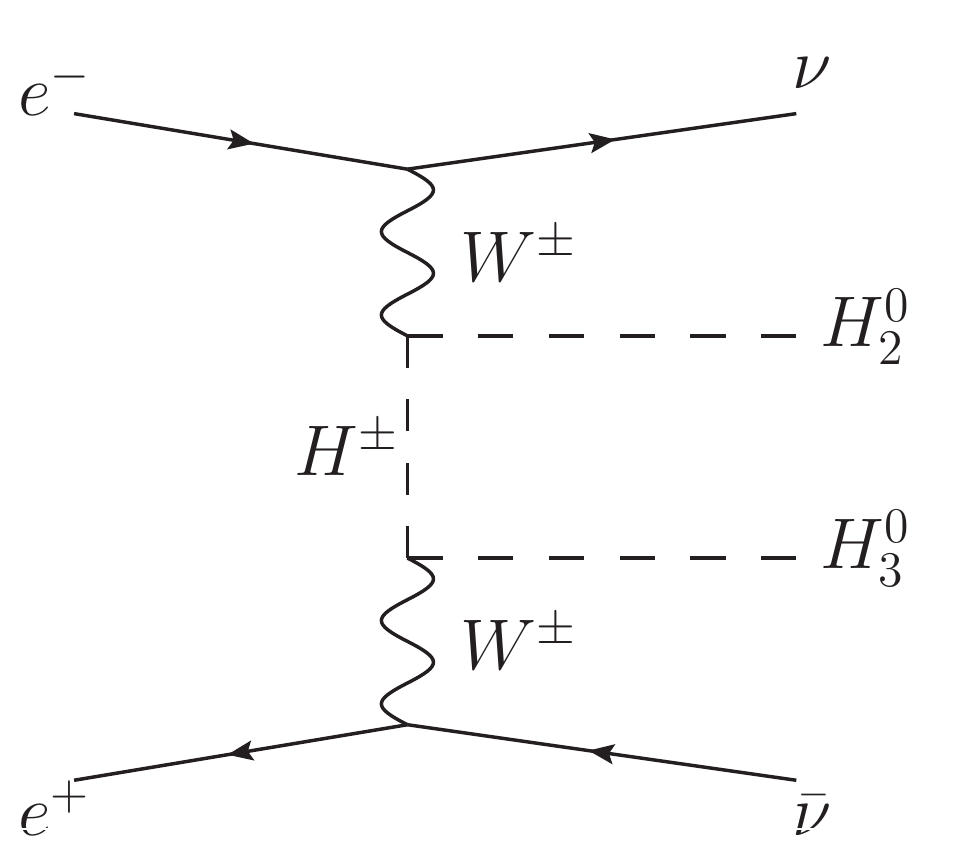}
		\caption{Typical Feynman diagrams of the pair production of the extra Higgs bosons.}
		\label{diagpairpro}
		\vspace{5 mm}
		\includegraphics[height=60 mm]{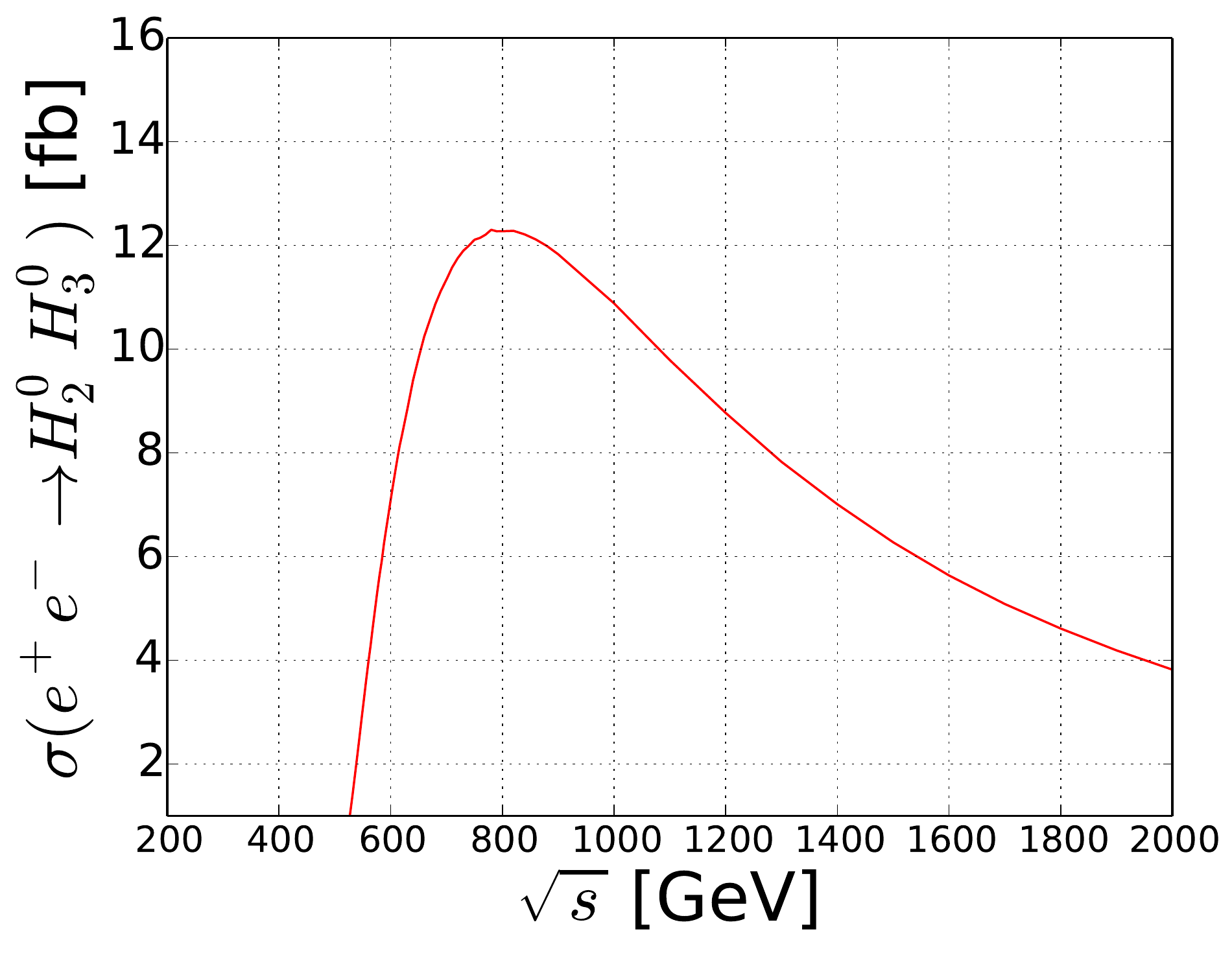}
		\includegraphics[height=60 mm]{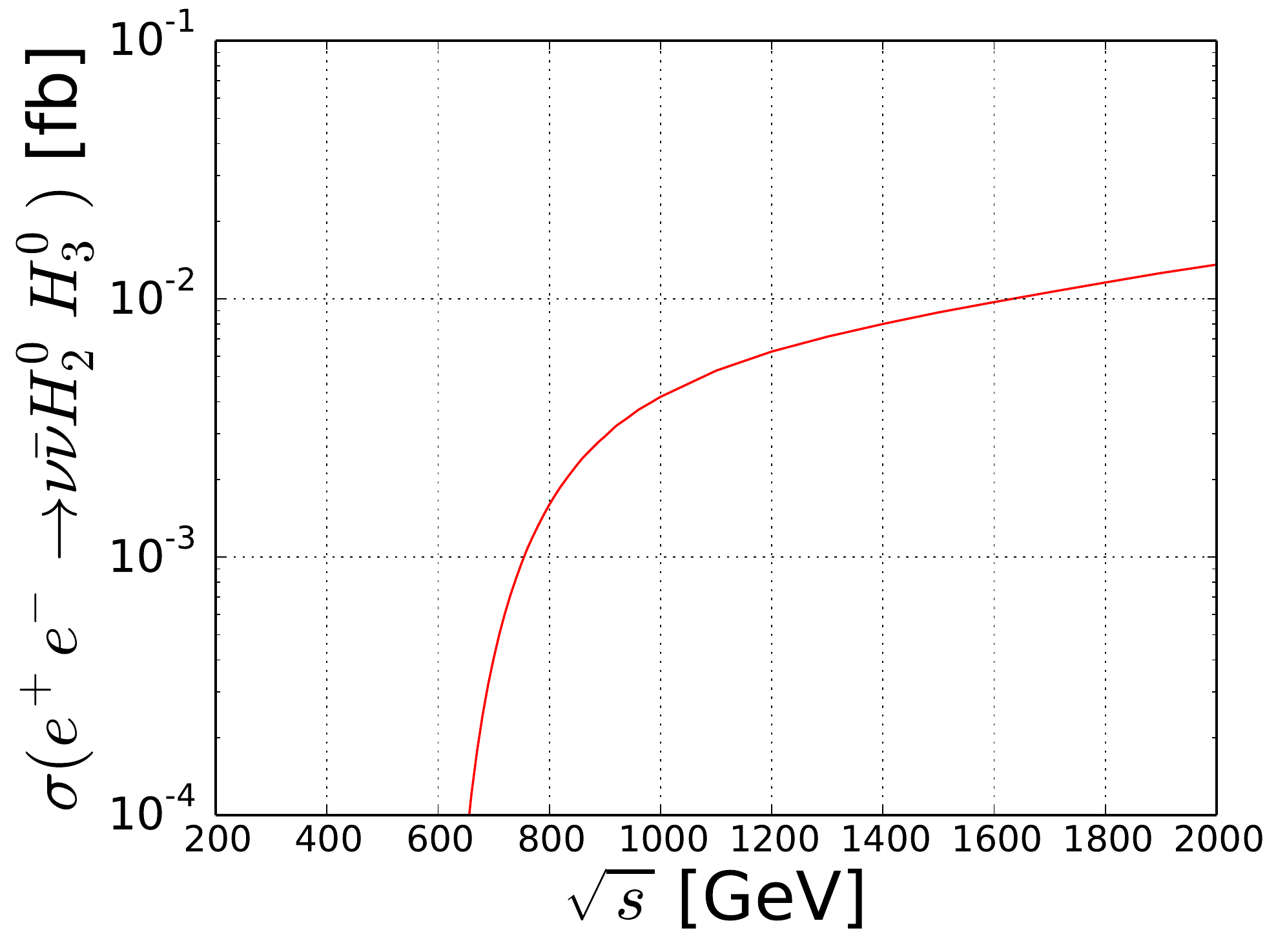}
		\caption{
			Cross sections of $e^+ e^- \to H^0_2 H^0_3$ (left) and $e^+ e^- \to \nu \bar{\nu} H^0_2 H^0_3$ (right) as a function of $\sqrt{s}$ for the benchmark point given in Tab.~\ref{tb:THDMinput}. 
			}
		\label{xsecpairproroots}
	\end{figure}

In our scenario, the additional Higgs bosons are mainly produced from the following processes,
\begin{align}
&e^+ e^- \to H^0_2 H^0_3,\\
&e^+ e^- \to \nu \bar{\nu} H^0_2 H^0_3,
\end{align}
whose diagrams are shown in Fig.~\ref{diagpairpro}. 
We note that the couplings of $H^0_jVV$ ($j=2,3$) vanish due to the alignment of the Higgs potential, so that the single productions of the extra Higgs boson from the Higgs-strahlung process and the vector boson fusion cannot be used at tree level. 
Assuming the expected performance of the ILC\cite{Baer:2013cma,Fujii:2015jha,Fujii:2017vwa}, we take the beam polarization for electrons (positrons) to be $-80$\% ($+30$\%)\footnote{The beam polarization is expressed as $(N_R-N_L)/(N_R+N_L)$, where $N_R$ and $N_L$ are number of the right-handed and left-handed particles, respectively\cite{Fujii:2018mli}. }, by which the cross section of $e^+ e^-\to H^0_2 H^0_3$ is enhanced by 138\% as compared with the unpolarized case. 
In Fig.~\ref{xsecpairproroots}, the cross sections are shown as a function of the collision energy $\sqrt{s}$. 
Since the $e^+ e^-\to H^0_2 H^0_3$ process is s-channel, the cross section is maximized when $\sqrt{s}$ is taken to be just above the threshold. 
In the benchmark point, the maximal value of the cross section is given to be about $12$ fb at $\sqrt{s}=800$ GeV\footnote{In this paper, we perform our analysis at the energy upgraded version of the ILC. }. 
It is seen that the cross section of $e^+ e^- \to \nu \bar{\nu} H^0_2 H^0_3$ is around three orders of magnitude smaller than that of $e^+ e^- \to H^0_2 H^0_3$. 
Therefore, we focus on $e^+ e^- \to H^0_2 H^0_3$ as the promising process to test the CP violation. 

\subsection{Signal and Background Processes}
By taking into account the decay property of the additional Higgs bosons and the reconstruction of the azimuthal angle $\Delta\phi$ discussed in previous section, the $e^+e^-\to H^0_2H^0_3\to b\bar{b}\tau^+\tau^-$ process is useful to test the CP-violating phase. 
For the signal process, we take the benchmark point given in Tab.~\ref{tb:THDMinput}. 
The decay rates of $H^0_2(H^0_3)\to\tau^+\tau^-$ and $H^0_2(H^0_3)\to b\bar{b}$ are then determined to be 11.8\% (54.2\%) and 9.92\% (45.5\%), respectively. 
The total decay width of $H^0_2(H^0_3)$ is given as $1.23\times10^{-3}$ ($2.20\times10^{-4}$) GeV. 
We take $\sqrt{s}=800$ GeV such that the signal cross section is maximized to be 12.3 fb and the integrated luminosity $\mathcal{L}=3000$ fb${}^{-1}$.

\begin{figure}
			\begin{tabular}{c}
				\begin{minipage}{0.5\hsize}
					\centering
					\includegraphics[width=70 mm]{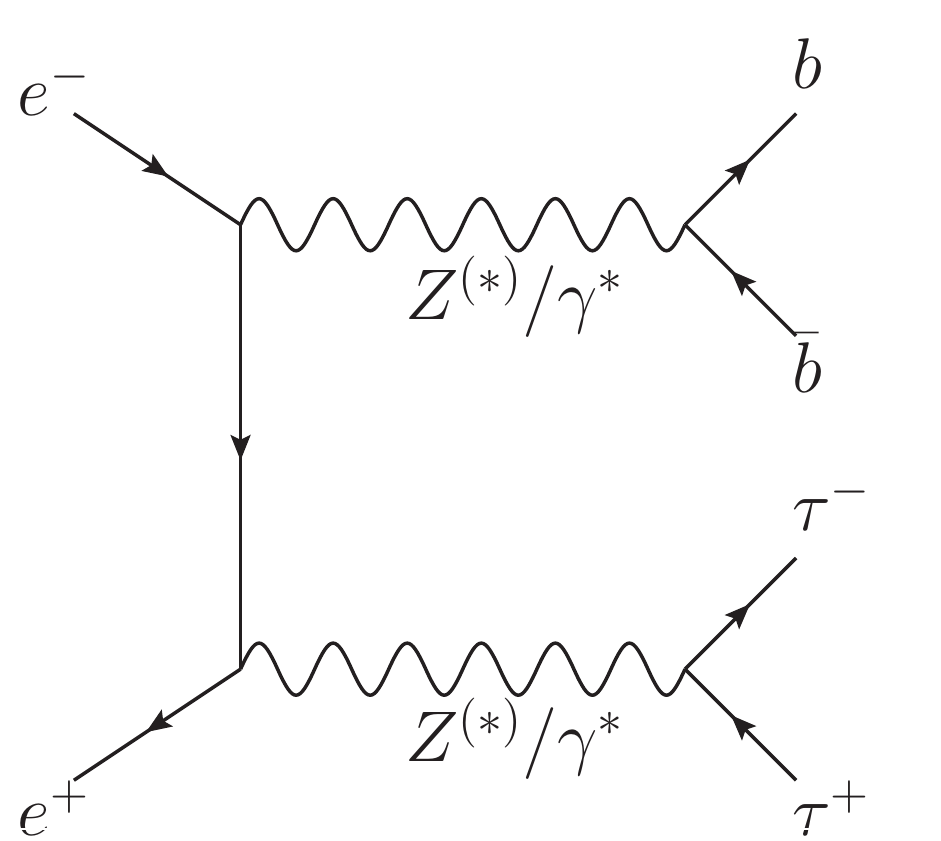}
					\\(\ref{fg:diagramsBG}-a)
				\end{minipage}
				%%%
				\begin{minipage}{0.5\hsize}
					\centering
					\includegraphics[width=70 mm]{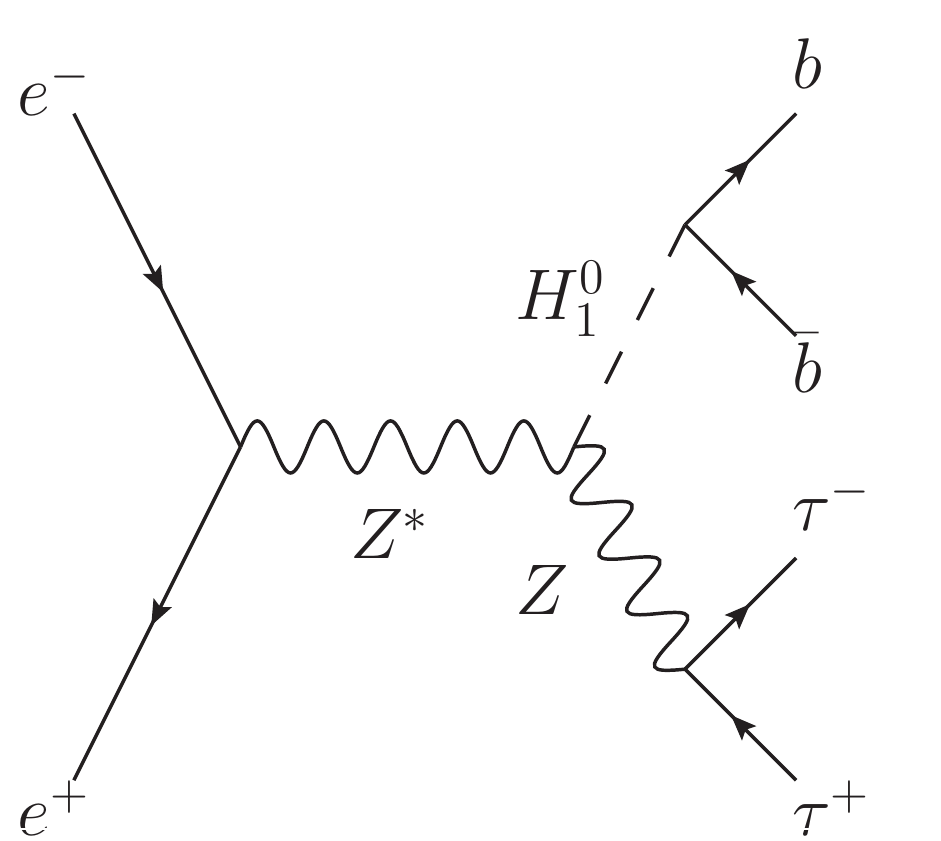}
					\\(\ref{fg:diagramsBG}-b)
				\end{minipage}
				%%%
				\\
				\begin{minipage}{0.5\hsize}
					\centering
					\includegraphics[width=70 mm]{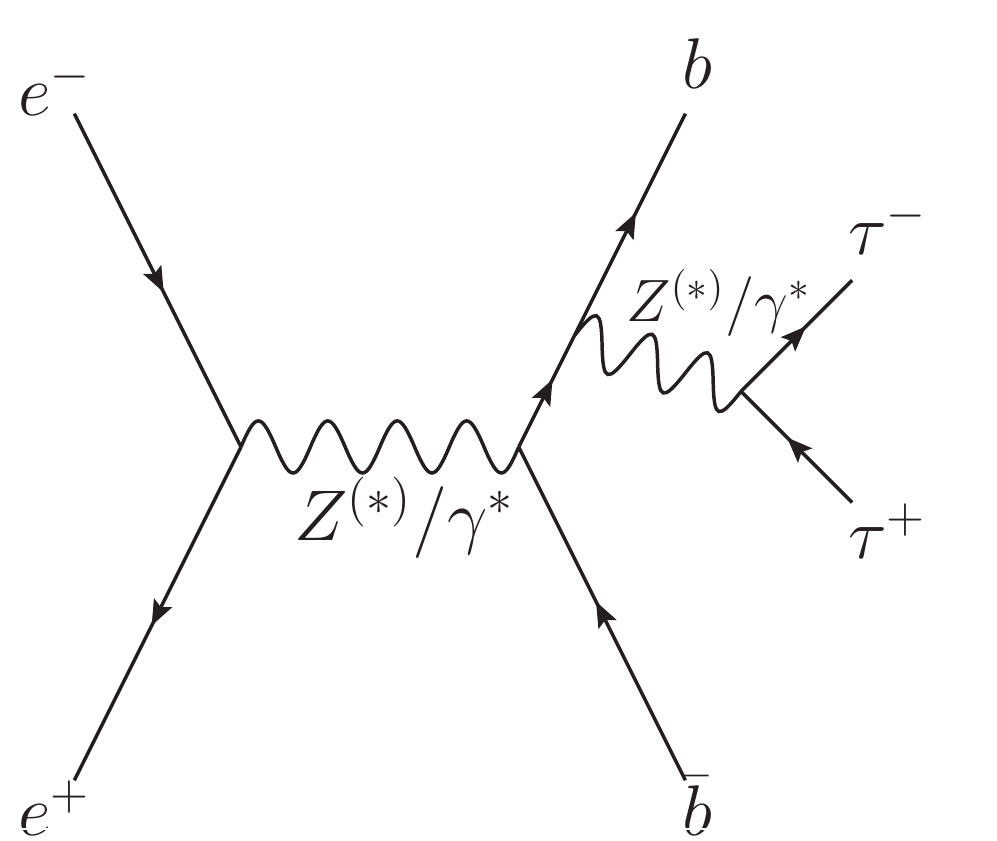}
					\\(\ref{fg:diagramsBG}-c)
				\end{minipage}
				%%%
				\begin{minipage}{0.5\hsize}
					\centering
					\includegraphics[width=70 mm]{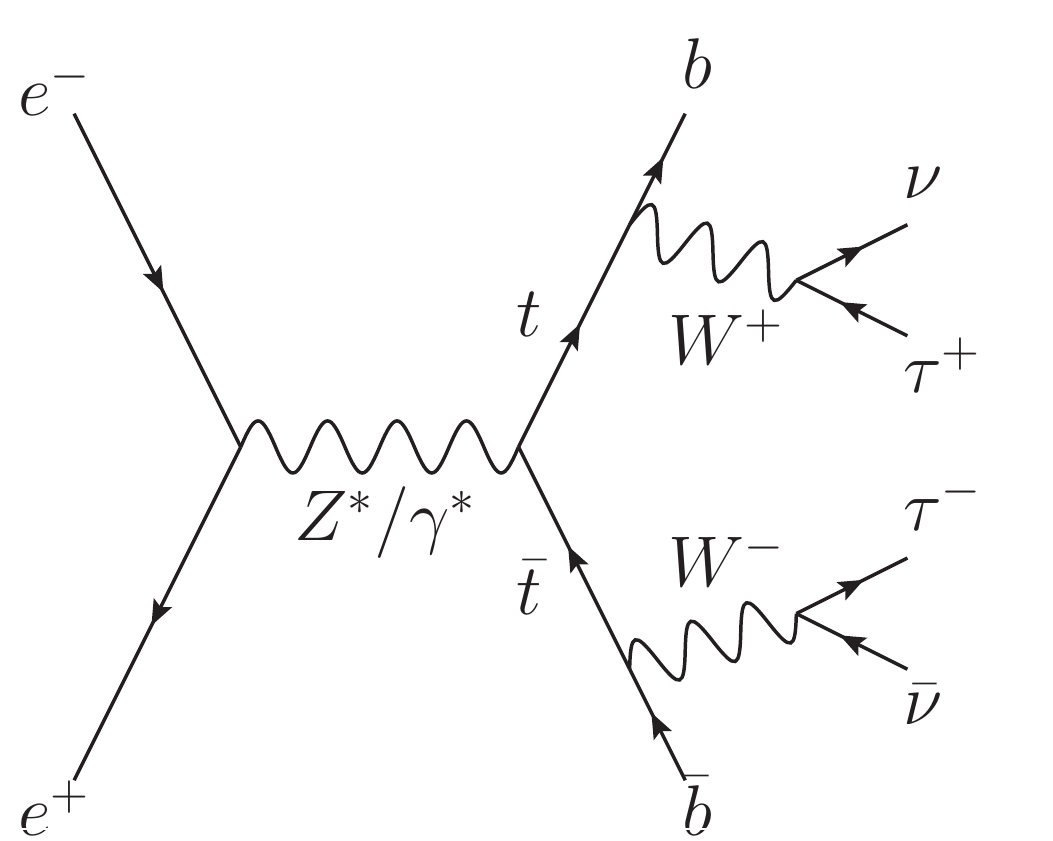}
					\\(\ref{fg:diagramsBG}-d)
				\end{minipage}
				%%%
			\end{tabular}
			\caption{Typical Feynman diagrams of the background processes.}
			\label{fg:diagramsBG}
		\end{figure}

We consider two types of the background processes. 
The first one has exactly the same final state as that of the signal process, which arises from the pair production of the neutral (off-shell) gauge bosons (Fig.~\ref{fg:diagramsBG}-a), the Higgs-strahlung process (Fig.~\ref{fg:diagramsBG}-b) and the Drell-Yan processes (Fig.~\ref{fg:diagramsBG}-c). 
The other one comes from the $t\bar{t}$ production with the top decay: $t\to bW\to b\tau\nu$ (Fig.~\ref{fg:diagramsBG}-d). 

We perform the simulation study by using MadGraph5\cite{Alwall:2011uj} for generations of $b\bar{b}\tau^+\tau^-(\nu\bar{\nu})$ events which are passed to TauDecay\cite{Hagiwara:2012vz} in order to consider the hadronic decays of the tau leptons, i.e., $\tau^\pm\to\pi^\pm\nu$, $\tau^\pm\to\rho^\pm\nu$ and $\tau^\pm\to a_1^\pm\nu$ with their subsequent decays $\rho^\pm\to\pi^\pm\pi^0$, $a_1^\pm\to\pi^\pm\pi^0\pi^0$ and $a_1^\pm\to\pi^\pm\pi^\pm\pi^\mp$. 
The branching ratios of these decays are given as 10.8\%($\tau^\pm\to\pi^\pm\nu$), 25.5\%($\tau^\pm\to\pi^\pm\pi^0\nu$), 9.26\%($\tau^\pm\to\pi^\pm\pi^0\pi^0\nu$) and 8.99\%($\tau^\pm\to\pi^\pm\pi^\pm\pi^\mp\nu$)\cite{Zyla:2020zbs}. 
We do not perform the detector level simulation. 
In addition, we assume the efficiency for the identification of a bottom quark jet to be 70\%, in which the mis-identification rate of jets from other quarks and gluons is estimated to be less than O(1)\%\cite{Suehara:2015ura}. 
For the decay mode of the tau leptons, we also take the efficiency for the identification of $\pi$, $\rho$ and $a_1$ mesons as 89.27\%, 75.21\% and 64.32\%, respectively\cite{Jeans:2019brt}. 
Although $a_1^\pm$ have mainly two decay modes: $a_1^\pm\to\pi^\pm\pi^0\pi^0$ and $a_1^\pm\to\pi^\pm\pi^\pm\pi^\mp$, we apply the same efficiency to both modes. 
Since the energy resolution of jets is expected to be $\Delta E=0.17\sqrt{E}$ at the ILC\cite{Behnke:2013lya}, the typical value of $\Delta E$ is given to be about 2 GeV for $E=\mathcal{O}(100)$ GeV. 
In the following analysis, we take a larger bin size such as 10 GeV for jet energies as a conservative choice.

\subsection{Distributions}
We show various distributions for the signal and background events. 

\begin{figure}
		\centering
		\includegraphics[width=90 mm]{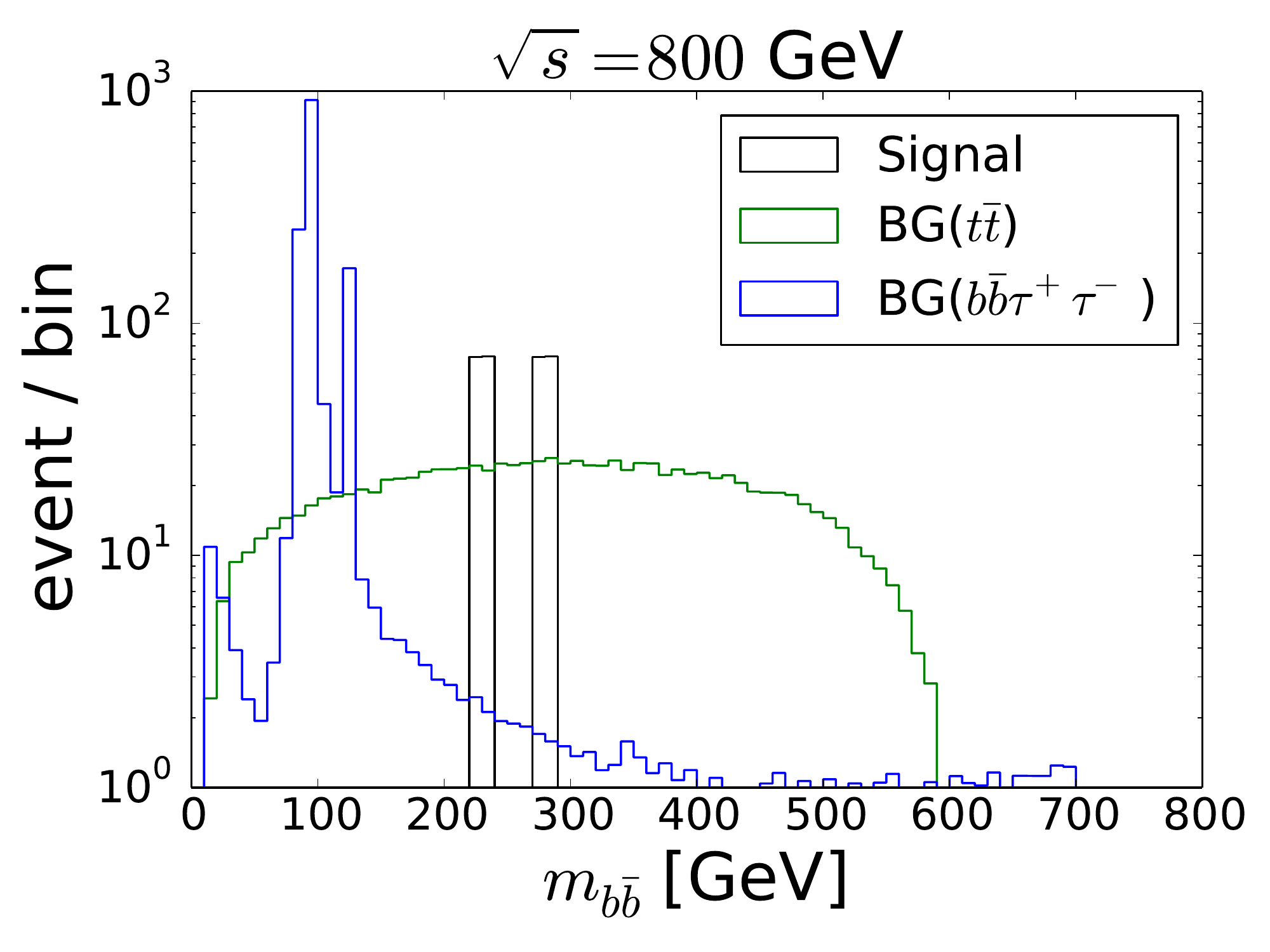}
		\caption{
			The $b\bar{b}$ invariant mass $m_{b\bar{b}}$ distribution for the benchmark point given in Tab.~\ref{tb:THDMinput} for $\sqrt{s}=800$ GeV and $\mathcal{L}=3000$ fb${}^{-1}$. 
			}
		\label{bbInvMass}
	\end{figure}

In Fig.~\ref{bbInvMass}, the $b\bar{b}$ invariant mass $m_{b\bar{b}}$ distributions for $e^+e^-\to b\bar{b}\tau^+\tau^-$ and $e^+e^-\to b\bar{b}\tau^+\tau^-\nu\bar{\nu}$ with the tau leptons decaying into $\pi$, $\rho$ and $a_1$ are shown. 
The black, green and blue histograms are the distributions given by the signal events, the background from the $t\bar{t}$ production and that from $b\bar{b}\tau^+\tau^-$, respectively.
We see the sharp peaks at around $m_{H^0_2}$ and $m_{H^0_3}$ ($m_Z$ and $m_{H^0_1}$) in the signal (background) events, while no particular structure is seen in the $t\bar{t}$ background. 
We thus impose the invariant mass cut $|m_{b\bar{b}}-m_{H^0_3}|\leq10$ GeV ($|m_{b\bar{b}}-m_{H^0_2}|\leq10$ GeV) which extracts the signal events containing the decay of $H^0_3$ ($H^0_2$) into $b\bar{b}$. 

\begin{figure}
		\centering
		\includegraphics[width=90 mm]{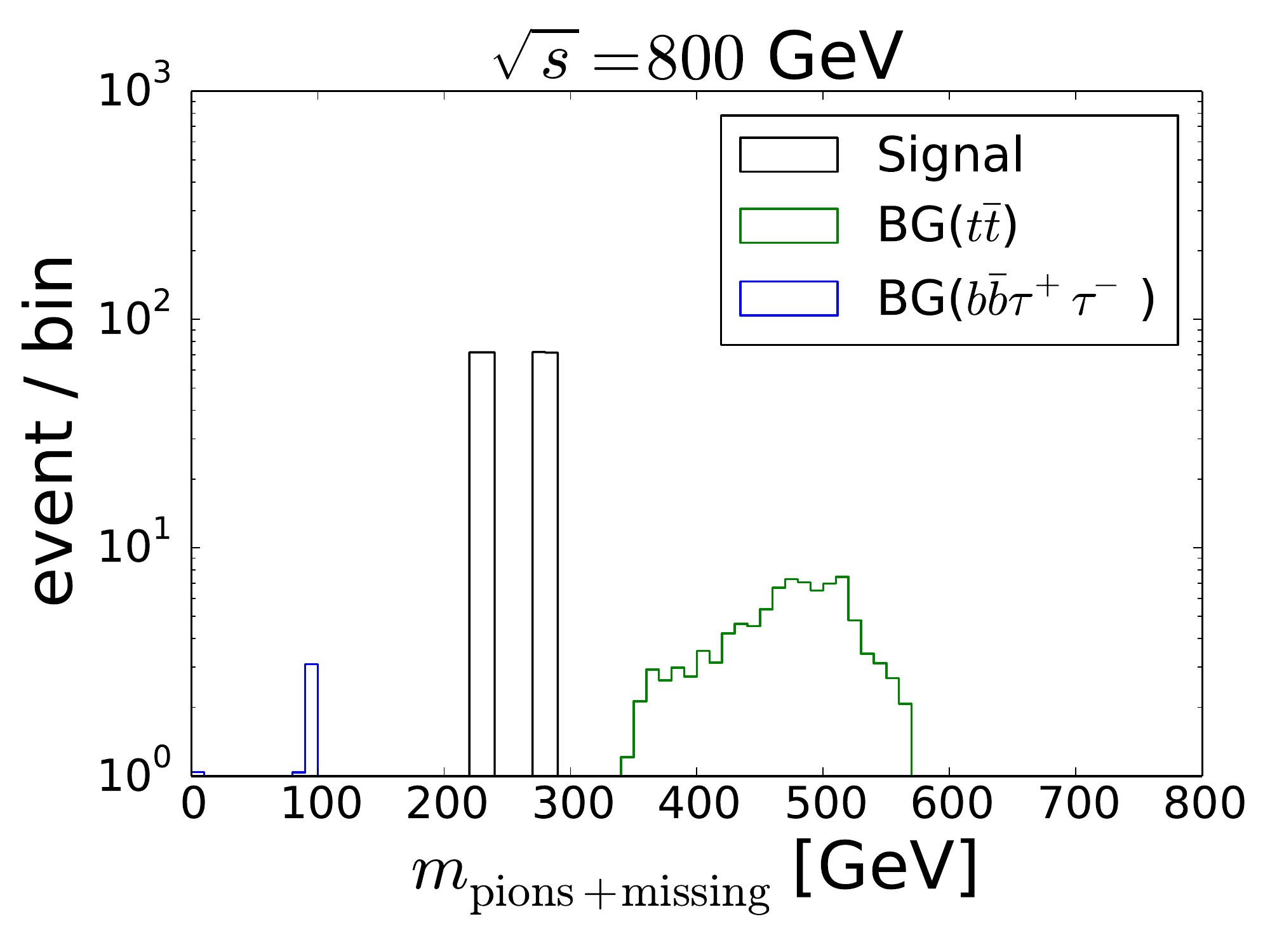}
		\caption{
			Invariant mass distribution of all pions produced by tau decay and missing momentum for the benchmark point given in Tab.~\ref{tb:THDMinput} after the cuts of the $b\bar{b}$ invariant mass such as $|m_{b\bar{b}}-m_{H^0_3}|\leq10$ GeV and $|m_{b\bar{b}}-m_{H^0_2}|\leq10$ GeV for $\sqrt{s}=800$ GeV and $\mathcal{L}=3000$ fb${}^{-1}$. 
			}
		\label{tautauInvMass}
	\end{figure}

In Fig.~\ref{tautauInvMass}, we show the invariant mass distribution $m_{\textrm{pions}+\textrm{missing}}$ for all the pions produced by the decays of tau leptons and missing momentum after applying the $m_{b\bar{b}}$ cuts. 
This invariant mass distribution can be reconstructed by all the visible particles of the final state and the information of the initial energy of the $e^+e^-$ collision. 
Again, we see the clear peaks at around $m_{H^0_2}$ and $m_{H^0_3}$ in the signal events, because $m_{\textrm{pions}+\textrm{missing}}$ corresponds to the invariant mass of the $\tau^+\tau^-$ system from the Higgs boson decay. 
We thus take further cut $|m_{\textrm{pions}+\textrm{missing}}-m_{H^0_2}|\leq10$ GeV ($|m_{\textrm{pions}+\textrm{missing}}-m_{H^0_3}|\leq10$ GeV), by which most of the background events can be removed. 

The number of events before and after applying the kinematic cuts for the signal and background processes is summarized in Tab.~\ref{tb:BM0}. 
After the two cuts of $m_{b\bar{b}}$ and $m_{\textrm{pions}+\textrm{missing}}$, no background event survives in our simulation. 
		\begin{table}
			\centering
			\begin{tabular}{|l||c|ccccc||c||c|} \hline
					& $t\bar{t}$ & $b\bar{b}\tau^+\tau^-$ & {\small $(ZZ)$} &{\small $(ZH^0_1)$} &{\small $(Z\gamma^*)$} & {\small $(\gamma^*\gamma^*)$} & {\small $H^0_2H^0_3$} & {\small $S/\sqrt{S+B}$} \\ \hline \hline
					No cut
						& 6093 & 10080 & (4205) & (1248) & (3662) & (179) & 1867 & 13.90 \\ \hline \hline
					{\small $|m_{b\bar{b}}-m_{H^0_{3/2}}|\leq10$ GeV}
						& 272/308 & 23/24 & (0/0) & (0/0) & (2/2) & (4/5) & 934/933 & 26.64/26.23 \\ \hline 
					{\small $|m_{\textrm{pions}+\textrm{missing}}-m_{H^0_{2/3}}|\leq10$ GeV}
						& 0/0 & 0/0 & (0/0) & (0/0) & (0/0) & (0/0) & 934/933 & 30.56/30.55 \\ \hline \hline
					Combined values after $m_{b\bar{b}}$ cut 
						& 580 & 47 & (0) & (0) & (4) & (9) & 1867 & 37.28 \\ \hline
					Combined values after two cuts
						& 0 & 0 & (0) & (0) & (0) & (0) & 1867 & 43.21 \\ \hline
			\end{tabular}
				\caption{
					Number of events of the signal and background of $e^+e^-\to b\bar{b}\tau^+\tau^-$ and $e^+e^-\to b\bar{b}\tau^+\tau^-\nu\bar{\nu}$ for the benchmark point given in Tab.~\ref{tb:THDMinput} with $\sqrt{s}=800$ GeV and $\mathcal{L}=3000$ fb${}^{-1}$. 
					The efficiency for the identification of a bottom quark jet is assumed to be 70\%. 
					The values in parentheses are the number of partial events of $e^+e^-\to b\bar{b}\tau^+\tau^-$. 
					In the second and third rows, the numbers given in the left (right) side of the slash are the number after imposing the invariant mass cut for $m_{H^0_2}$ and $m_{H^0_3}$. 
					The numbers given in the last two rows show the combined values obtained from the different cuts for $m_{H^0_2}$ and $m_{H^0_3}$. 
					}
				\label{tb:BM0}
		\end{table}

\begin{figure}
		\centering
		\includegraphics[width=90 mm]{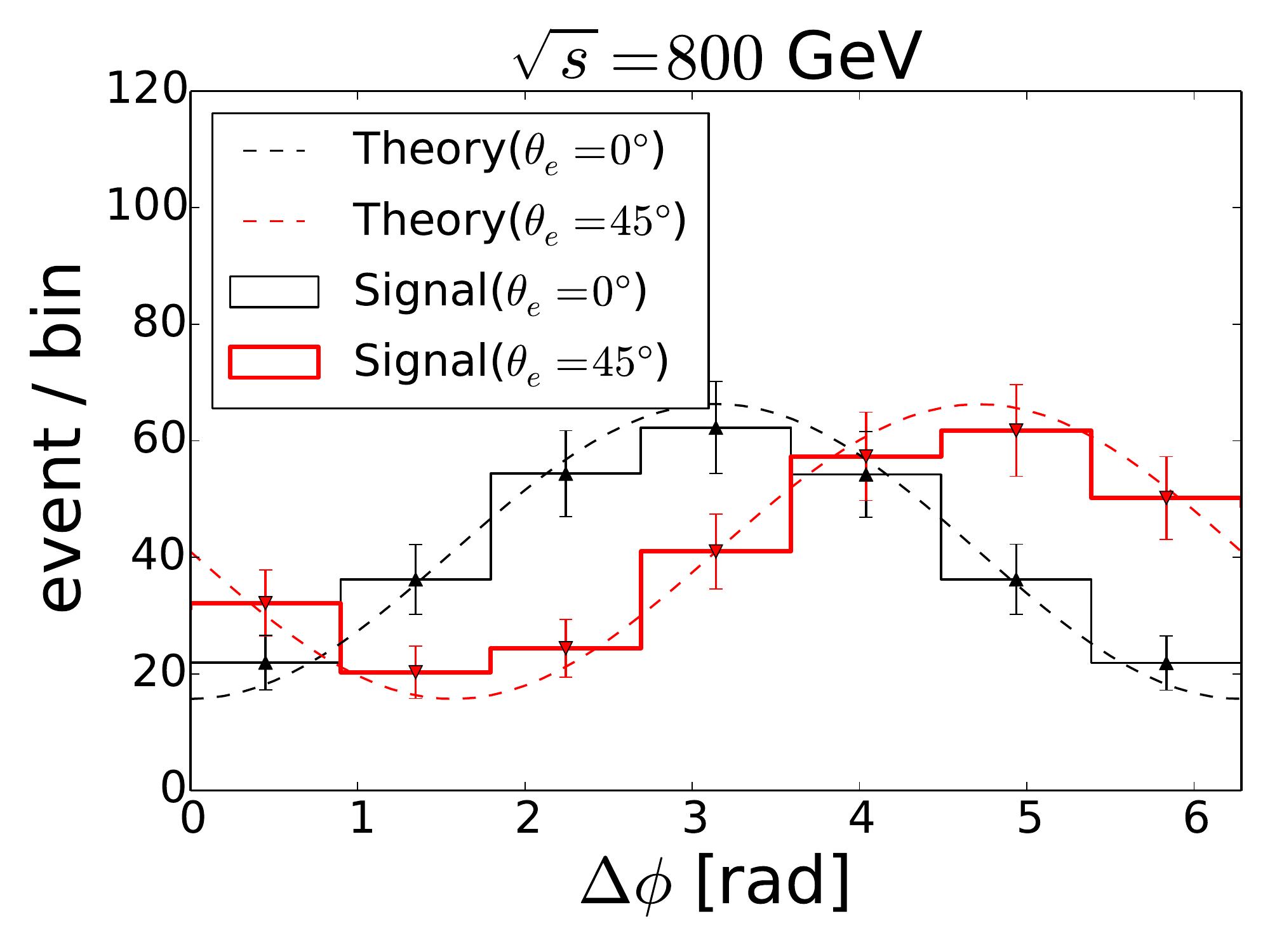}
		\caption{
			The $\Delta\phi$ distribution for the benchmark in Tab.~\ref{tb:THDMinput} for $0\leq\Delta\phi<2\pi$ for $m_{H^0_2}=280$ GeV, $m_{H^0_3}=m_{H^\pm}=230$ GeV, $\sqrt{s}=800$ GeV and $\mathcal{L}=3000$ fb${}^{-1}$. 
			The signal events are compared with the theoretical prediction (dashed line). 
			The bars for each of the bins correspond to the 1$\sigma$ error. 
			For the seventh bin of $\Delta\phi\sim$ 5.4-6.3, the 6.1$\sigma$ error of the black histogram ($\theta_e=0$) cannot reach the red histogram ($\theta_e=\pi/4$). 			}
		\label{angledistBM0}
	\end{figure}

In Fig.~\ref{angledistBM0}, we show the $\Delta\phi$ distribution after applying the $m_{b\bar{b}}$ and $m_{\textrm{pions}+\textrm{missing}}$ cuts. 
For the signal distribution, we combine the events from $H^0_2\to\tau^+\tau^-$, $H^0_3\to b\bar{b}$ and those from $H^0_2\to b\bar{b}$, $H^0_3\to\tau^+\tau^-$ by shifting the latter distribution with $\pi$ radian, which can be justified due to our assumption, i.e., the phases of the Yukawa couplings of $H^0_2$ and $H^0_3$ are different by $\pi/2$ radian. 
The red and black lines correspond to the scenario with $\theta_e=\pi/4$ and $0$, and the bars for each of the bins correspond to the 1$\sigma$ error. 
From this figure, at the third (seventh) bin from left in which $\Delta\phi$ is about 1.8-2.7 (5.4-6.3), it is seen that the data of the red histogram is $-4.0$ ($+6.1$) $\sigma$ away from these of the black histogram. 

Next, we survey the parameter region around the benchmark point. 
We focus on the seventh bin of the $\Delta\phi$ distribution, and define the significance as
	%%%
		\eq{
			\Delta_\textrm{CP}\equiv(N_{\theta_e=\pi/4}-N_\textrm{CPC})/\sqrt{N_\textrm{CPC}}
		,}
where $N_{\theta_e=\pi/4}$ ($N_\textrm{CPC}$) is the number of signal events with $\theta_e=\pi/4$ (without CP-violating phases) in the seventh bin. 
The number of signal events for any scenarios are estimated from these of the benchmark point by the production cross section and the branching ratios of the additional Higgs bosons. 
We note that the stability of the EDM cancellation at a high energy is not always realized for the scenarios different from the benchmark point.

We perform the scan for the coupling strength $|\zeta_e|\in(0,2]$ and the additional Higgs masses $m_{H^0_2}\in(100,500]$ with the fixed mass difference $m_{H^0_3}=m_{H^0_2}-50$ GeV. 
The other parameters are taken to be the same as the benchmark point. 
On the other hand, we also investigate the excluded region against for the electron EDM data with $|\lambda_7|=0.01$, $0.1$, $0.3$, $0.5$ and $0.7$ under the scan of $\theta_7\in(-\pi,\pi]$, where $\theta_7$ and $|\lambda_7|$ do not affect on the collider phenomenology discussed above.

\begin{figure}
			\begin{tabular}{c}
				\begin{minipage}{0.5\hsize}
					\centering
					\includegraphics[width=80 mm]{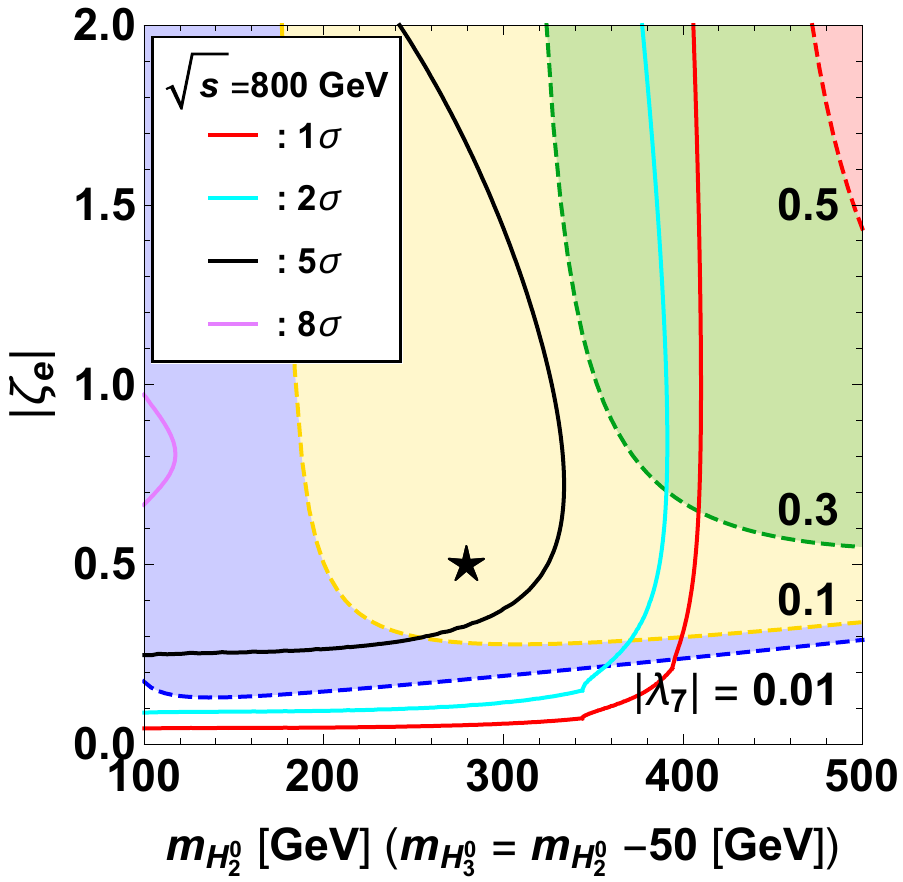}
					%\\(\ref{fg:}-a)
				\end{minipage}
				%%%
				\begin{minipage}{0.5\hsize}
					\centering
					\includegraphics[width=80 mm]{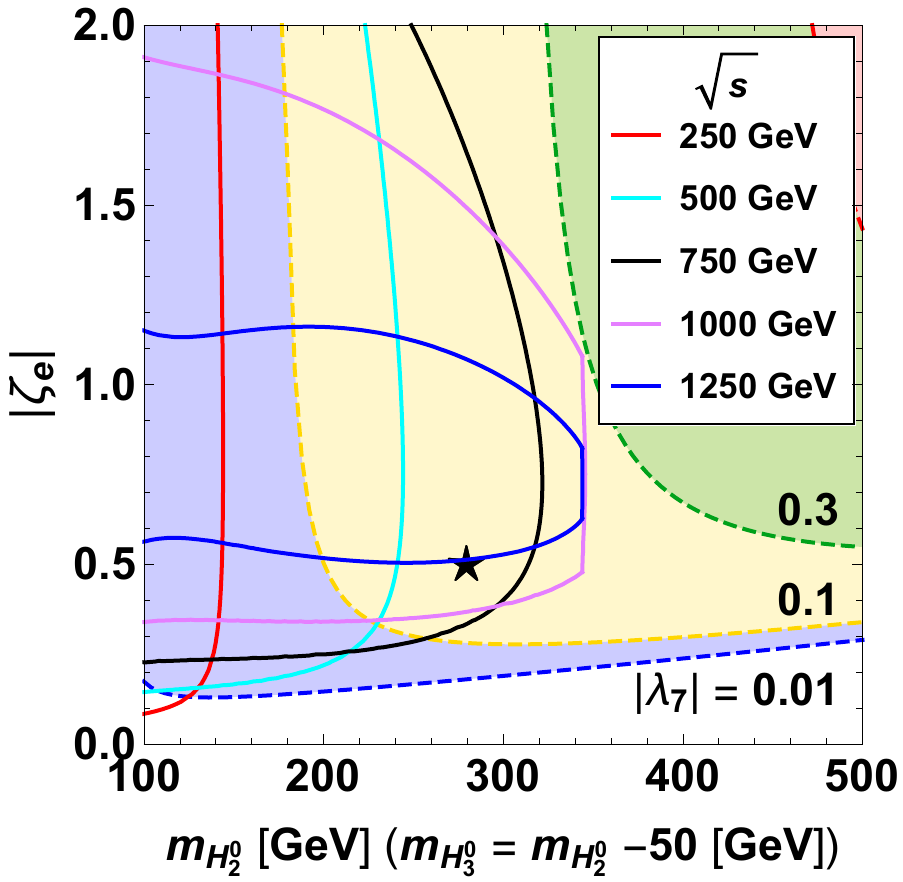}
					%\\(\ref{fg:}-b)
				\end{minipage}
				%%%
			\end{tabular}
		\caption{
			(Left) Contour plot of the significance $\Delta_\textrm{CP}$ (red: $\Delta_\textrm{CP}=1$, cyan: $\Delta_\textrm{CP}=2$, black: $\Delta_\textrm{CP}=5$ and purple: $\Delta_\textrm{CP}=8$) on the $m_{H^0_2}$-$|\zeta_e|$ plane in the case with $|\zeta_u|=0.01$, $|\zeta_d|=0.1$, $\theta_u=1.2$, $\theta_d=0$, $\theta_e=\pi/4$, $m_{H^\pm}=m_{H^0_3}$ and $m_{H^0_3}=m_{H^0_2}-50$ GeV. 
			The blue, yellow, green and red shaded regions (the regions above each dashed curve) are excluded by the electron EDM data for $|\lambda_7|=0.01$, $0.1$, $0.3$ and $0.5$, respectively, under the scan of $\theta_7\in(-\pi,\pi]$. 
			The point marked by the star corresponds to the benchmark point when $\lam_7=0.3$ and $\theta_7=-1.8$. 
			(Right) The points on the solid curves give $\Delta_\textrm{CP}=5$ with a fixed value of $\sqrt{s}$ to be $250$ (red), $500$ (cyan), $750$ (black), $1000$ (purple) and $1250$ (blue) GeV. 
			The other configuration is the same as the left panel. 
			}
		\label{SigExcl}
	\end{figure}

In the left panel of Fig.~\ref{SigExcl}, the contour plot of the significance $\Delta_\textrm{CP}$ is denoted by the solid lines. 
The excluded regions against for the electron EDM data are denoted by the color shaded regions above the dashed lines. 
The point marked by the star corresponds to the benchmark point when $\lam_7=0.3$ and $\theta_7=-1.8$. 
In the area of the left side of the red (cyan, black and purple) solid line, we can get more than 1 (2, 5 and 8) $\sigma$ significance on the $\Delta\phi$ distribution for $\sqrt{s}=800$ GeV. 
It is seen that the significance is enhanced in the parameter space with smaller masses and about $|\zeta_e|=0.5$ because the number of the evens of $e^+e^-\to\tau^+\tau^- b \bar{b}$ increases. 
When the additional Higgs bosons are heavier than $2m_t$, the $H^0_j\to t\bar{t}$ cannel opens, and the significance is drastically reduced due to the small branching ratios of $H^0_{2,3}\to b\bar{b}/\tau^+\tau^-$. 
The color shaded region above the blue (yellow, green and red) dashed line denotes the excluded region against for the electron EDM data which are not allowed even if we take any value of $\theta_7\in(-\pi,\pi]$ when $|\lam_7|$ are fixed as $0.01$ ($0.1$, $0.3$ and $0.5$). 
The color shaded regions are piled up in the order of blue, yellow, green and red. 
It is seen that the cancellation works well for the wide regions for $|\zeta_e|$ in which the electron EDM is satisfied, when $|\lam_7|=0.1$ (0.3, 0.5) and $m_{H^0_2}$ is smaller than about 200 (320, 460) GeV. 
On the other hand, if the coupling strength $|\lam_7|$ is smaller or the masses are larger, the cancellation does not work due to the unbalance of the two contributions between the fermion-loops and the Higgs boson-loops on the BZ diagrams, so that the allowed regions remain in which the small value of the overall factor $|\zeta_e|$ on the electron EDM. 
We note that if the electron EDM bound is improved by one order of magnitude, the heights of all the dashed lines are made about 1/10, so that the parameter regions in which the cancellation does not work well are surveyed. 
It is seen the remaining parameter regions in which the cancellation works well can be tested from the $\Delta\phi$ distribution. 

In the right panel of Fig.~\ref{SigExcl}, the solid curves show the points giving $\Delta_\textrm{CP}=5$ with a fixed value of $\sqrt{s}=250$, $500$, $750$, $1000$ and $1250$ GeV. 
At the point marked by the star, the parameters are taken to be the same as the benchmark point when $\lam_7=0.3$ and $\theta_7=-1.8$. 
The color shaded regions are the same as those in Fig.~\ref{SigExcl}. 
Since the decay cannel $H^0_j\to t\bar{t}$ ($j=2,3$) opens when the additional Higgs bosons have masses larger than $2m_t$, the significance quickly decreases due to the dominance of this cannel on the decay of $H^0_j$. 
For $m_{H^0_{2,3}}<2m_t$, the test of CP violation by using the $\Delta\phi$ distributions can play a complement role to the future EDM experiments when $\lam_7\leq3$. 
From the above analysis, the CP-violating scenario can be distinguished from the CP-conserving case at the ILC.

Finally, we mention the testability of the CP property in our scenario at the LHC. 
In the aligned scenario, the neutral Higgs bosons are produced from the gluon-fusion process $gg\to H^0_{2, 3}$ and the bottom quark associated production $gg\to H^0_{2, 3}b\bar{b}$. 
However, these production cross sections are very small, typically a few fb level in our benchmark point, due to the small values of the Yukawa couplings, i.e., $\zeta_u=0.01$ and $\theta_d=0.1$\cite{Kanemura:2020ibp}. 
There are pair production processes such as $pp\to H^0_2H^0_3$, $H^0_{2 ,3}H^\pm$\cite{Cao:2003tr,Kanemura:2001hz,Belyaev:2006rf} and $H^+H^-$, and their cross sections are $\mathcal{O}(10)$ fb for the mass of the extra Higgs bosons to be $\mathcal{O}(200)$ GeV\cite{Kanemura:2020ibp}. 
On the other hand, the cross section of the $t\bar{t}$ production which would be the main background are measured to be about $10^{3}$ pb at the LHC with $\sqrt{s}=13$ TeV\cite{Aad:2020tmz}. 
Thus, at the LHC, it is quite challenging to get the enough significance to see the $\Delta\phi$ distribution. 
In addition, since initial energies for partons cannot be known, the reconstruction method for the tau leptons used in our analysis cannot be simply applied.

%%%%%%%%%%     discussions and conclusions     %%%%%%%%%%
\section{Discussions and Conclusions}\label{sc:summary}
We have discussed the the CP-violating THDM in which the Yukawa alignment is assumed to avoid the FCNCs at tree level and the alignment in the Higgs potential is also assumed to realize that the coupling constants for the 125 GeV Higgs boson are the same as those of the SM at tree level. 
In Ref.~\cite{Kanemura:2020ibp}, we have found the non-trivial parameter regions in which the electron EDM can be satisfied by considering the destructive interference between the Barr-Zee type contributions even if the CP-violating phases have O(1). 

In this work, for such a EDM suppressing scenario with O(1) phases, we have investigated how to test the CP violation at future lepton colliders like the ILC. 
In our scenario, since the alignment in the Higgs potential is imposed, the phenomenology of the extra Higgs bosons is important. 
In particular, we have considered the processes containing the decays of the extra neutral Higgs bosons into a pair of the tau leptons and we have performed the simulation in order to know the feasibility to extract the information of the CP-violating phase. 
From the signal and background simulation, we have found that the scenario with finite $\theta_e$ may be distinguished from the CP-conserving scenario at the energy upgraded version of the ILC with the integrated luminosity of 3000 fb${}^{-1}$.

In our scenario, we have mainly considered relatively smaller mass differences among the extra Higgs bosons. 
For larger mass differences, Higgs to Higgs decay modes can also be used to test the CP-violating effect as discussed in the recent paper\cite{Low:2020iua}. 

As we have mentioned in Ref.~\cite{Kanemura:2020ibp}, this scenario with O(1) phases have the possibility of electroweak baryogenesis. 
According to the discussion in Ref.~\cite{Cline:2011mm}, even if the vacuum expectation values of the Higgs doublets are taken to be $(\eval{\Phi_1^0},\eval{\Phi_2^0})=(v/\sqrt{2},0)$ at the zero temperature,
they can be $(\eval{\Phi_1^0},\eval{\Phi_2^0})=(v_1',v_2')/\sqrt{2}$ at the finite temperature. 
If such a phase transition can be realized, the top quark mass can be complex via $\zeta_u$ in the process of the strongly first order electroweak phase transition, and the BAU may be then generated. 
In order to know whether the BAU can be explained in our scenario while the current EDM data are satisfied at the same time, we have to perform the analysis for the electroweak phase transition and baryon number creation at the finite temperature. 
We also comment on the scenario of the loss of the alignment on the scalar potential at finite temperature but the alignment occurs at the zero temperature. 
In this case, both vacuum expectation values for both the scalar doublets can be non-zero at the finite temperature. 
It is important to survey the effect to BAU in such a scenario, which is performed elsewhere.

%%%%%%%%%%     acknowledgments     %%%%%%%%%%
\begin{acknowledgments}
We are grateful to Dr. Tanmoy Modak for a fruitful discussion at the first stage of this work. 
The work of S. K. was supported in part by Grant-in-Aid for Scientific Research on Innovative Areas, the Ministry of Education, Culture, Sports, Science and Technology, No.~16H06492 and also by JSPS, Grant-in-Aid for Scientific Research, Grant No. 18F18022, No.~18F18321 and No.~20H00160. 
The work of K. Y. was supported in part by the Grant-in-Aid for Early-Career Scientists, No.~19K14714. 
\end{acknowledgments}

%%%%%%%%%%     appendix     %%%%%%%%%%
\newpage
\appendix

\section{Exact Formulae of the Barr-Zee Type Contributions}\label{sc:BZformulae}
We here present the Barr-Zee type contributions to the EDM (CEDM) for a fermion (quark) $d_f^\gamma$, $d_f^Z$ and $d_f^W$ $(d_q^C)$ given in Eqs.~\eqref{eq:BZcontribution1} and \eqref{eq:BZcontribution2} in Sec.~\ref{sc:EDM}. 
The fermion-loop contributions to the EDM are
	%%%   fermion-loop BZ
		\small
		\begin{align}
				d_f^V(f')	&=\frac{em_f}{(16\pi^2)^2}8g_{Vff}^vg_{Vf'f'}^v
											Q_{f'}N_C\frac{m_{f'}^2}{v^2}
									\nn\\&\quad\quad
									\times\sum_j^3\int_0^1dz\Bigg\{
												\im[\kappa_f^j]\re[\kappa_{f'}^j]
												\rbra{\frac{1}{z}-2(1-z)}
												+\re[\kappa_f^j]\im[\kappa_{f'}^j]
												\frac{1}{z}
											\Bigg\}
									C^{VH^0_j}_{f'f'}(z)
				,\\
				d_{f(I_f=-\frac{1}{2})}^W(tb)&=\frac{eg_2^2m_f}{(16\pi^2)^2}N_C%|V_{tb}|^2
									\int_0^1dz\Bigg\{
											\frac{m_t^2}{v^2}\im[\zeta_f^*\zeta_u]\frac{2-z}{z}
											+\frac{m_b^2}{v^2}\im[\zeta_f^*\zeta_d]
									\Bigg\}[Q_t(1-z)+Q_bz]
									C^{WH^\pm}_{tb}(z)
				,\\
				d_{f(I_f=+\frac{1}{2})}^W(tb)&=\frac{eg_2^2m_f}{(16\pi^2)^2}N_C%|V_{tb}|^2
									\int_0^1dz\Bigg\{
											\frac{m_t^2}{v^2}\im[\zeta_f\zeta_u^*]
											+\frac{m_b^2}{v^2}\im[\zeta_f\zeta_d^*]\frac{1+z}{1-z}
									\Bigg\}[Q_t(1-z)+Q_bz]
									C^{WH^\pm}_{tb}(z)
		,\end{align}
		\normalsize
the Higgs boson-loop contributions to EDM are
	%%%   scalar-loop BZ
		\begin{align}
				d_f^V(H^\pm)		&=\frac{em_f}{(16\pi^2)^2}4g_{Vff}^v(ig_{H^+H^-V})
									\sum_j^3\im[\kappa_f^j]\frac{g_{H^\pm H^\mp H^0_j}}{v}
										\int_0^1dz(1-z)
									C^{VH^0_j}_{H^{^\pm}H^{^\pm}}(z)
				,\\
				d_f^W(H^\pm H^0)		&=\frac{eg_2^2m_f}{(16\pi^2)^2}\frac{1}{2}
									\sum_j^3\im[\kappa_f^j]\frac{g_{H^{^\pm} H^{^\mp} H^0_j}}{v}
										\int_0^1dz(1-z)
										C^{WH^\pm}_{H^{^\pm}H^0_j}(z)
		,\end{align}
the gauge-loop contributions to EDM are
	%%%   gauge-loop BZ
		\small
		\begin{align}
				d_f^V(W)		&=\frac{em_f}{(16\pi^2)^2}8g_{Vff}^v(ig_{WWV})\frac{m_W^2}{v^2}
									\sum_j^3\mathcal{R}_{1j}\im[\kappa_f^j]
									\nn\\&\quad\quad
										\times\int_0^1dz\Bigg[
											\cbra{
												\rbra{6-\frac{m_V^2}{m_W^2}}
												+\rbra{1-\frac{m_V^2}{2m_W^2}}\frac{m_{H^0_j}^2}{m_W^2}
											}\frac{(1-z)}{2}
											-\rbra{4-\frac{m_V^2}{m_W^2}}\frac{1}{z}
										\Bigg]
									C^{VH^0_j}_{WW}(z)
				,\\
				d_f^W(WH^0)	&=\frac{eg_2^2m_f}{(16\pi^2)^2}\frac{1}{2}\frac{m_W^2}{v^2}
									\sum_j^3\mathcal{R}_{1j}\im[\kappa_f^j]
										\int_0^1dz\Bigg\{
										\frac{4-z}{z}-\frac{m_{H^\pm}^2-m_{H^0_j}^2}{m_W^2}
										\Bigg\}(1-z)
									C^{WH^\pm}_{WH^0_j}(z)
		,\end{align}
		\normalsize
and the quark-loop contributions to CEDM are
	%%%   BZ CEDM
		\small
		\begin{align}
				d_q^C(q')	&=\frac{m_q}{(16\pi^2)^2}4g_3^3
											\frac{m_{q'}^2}{v^2}
									\sum_j^3\int_0^1dz\Bigg\{
												\im[\kappa_q^j]\re[\kappa_{q'}^j]
												\rbra{\frac{1}{z}-2(1-z)}
												+\re[\kappa_q^j]\im[\kappa_{q'}^j]
												\frac{1}{z}
											\Bigg\}
									C^{gH^0_j}_{q'q'}(z)
		,\end{align}
		\normalsize
where $V=\gamma$, $Z$ and $N_C$ is the color factor. 
The coupling constants are given by
$g_{\gamma ff}^v=eQ_f$,
$g_{Z ff}^v=g_z(I_f/2-Q_fs_W^2)$,
$g_{H^+ H^- \gamma}=-ie$,
$g_{H^+ H^- Z}=-ig_Z c_{2W}/2$,
$g_{H^\pm H^\mp H^0_j}=(\lam_3\mathcal{R}_{1j}+\re[\lam_7]\mathcal{R}_{2j}-\im[\lam_7]\mathcal{R}_{3j})v$,
$g_{WWA}=-ie$ and
$g_{WWZ}=-ig_Zc_W^2$,
where $Q_f$ are the electric charges of the fermion, $g_Z=\sqrt{g_1^2+g_2^2}$, $s_W=\sin\theta_W$, $c_W=\cos\theta_W$ and $c_{2W}=\cos2\theta_W$. 
	%%%
		\begin{align}
				C^{GH}_{XY}(z)	&=C_0\rbra{0,0;m_G^2,m_H^2,\frac{(1-z)m^2_X+zm_Y^2}{z(1-z)}}
		,\end{align}
and where $C_0$ is the Passarino-Veltman function\cite{Passarino:1978jh},
	%%%
		\begin{align}
			C_0(0,0;m_1^2,m_2^2,m_3^2)=\frac{1}{m_1^2-m_2^2}
									\cbra{
										\frac{m_1^2}{m_1^2-m_3^2}\log\rbra{\frac{m_3^2}{m_1^2}}
										-\frac{m_2^2}{m_2^2-m_3^2}\log\rbra{\frac{m_3^2}{m_2^2}}
										}
		.\end{align}
We confirmed the consistency of our results with Refs.~\cite{Leigh:1990kf,BowserChao:1997bb,Jung:2013hka,Abe:2013qla,Cheung:2014oaa,Cheung:2020ugr}. 

\section{Decay of the Extra Higgs Bosons}\label{sc:decayrate}
We here present the partial decay width of the additional Higgs bosons discussed in Sec.~\ref{decay}.
We refer to Refs.~\cite{Bian:2017jpt,Djouadi:2005gi,Djouadi:2005gj,Christova:2002uja}. 
	%%%   partial decay width
		\begin{align}
			\Gamma(H^0_i\to f\bar{f})	
				&=	\frac{N_C G_{F} m_{H^0_i} m_f^2}{4\sqrt{2}\pi}
					\left(1-\frac{4m_f^2}{m_{H^0_i}^2}\right)^{3/2}
					\left[
						(\re[\kappa_f^i])^2
						+(\im[\kappa_f^i])^2\left(1-\frac{4m_f^2}{m_{H^0_i}^2}\right)^{-1}
					\right]
			,\\
			\Gamma(H^0_i\to gg)
				&=	\frac{G_F \alpha_s^2 (m_{H^0_i}) m_{H^0_i}^3}{64 \sqrt2 \pi^3}
					\left[
						\left| \sum_q \re[\kappa_q^i] A_{1/2}^{H} (\tau^i_q) \right|^2
						+\left| \sum_q \im[\kappa_q^i] A_{1/2}^{A} (\tau^i_q) \right|^2
					\right]
			,\\
			\Gamma(H^0_i \to \phi V^*)
				&=	\frac{9 G_\mu^2 m_V^4}{8 \pi^3}
					\delta_V' m_{H^0_i} g_{H^0_i \varphi V}^2
					\,G\rbra{\frac{m_\varphi^2}{m_{H^0_i}^2},\frac{m_V^2}{m_{H^0_i}^2}}
			,\\
			\Gamma(H^+\to ff')
				&=	\frac{N_C}{16\pi m_{H\pm}^3}
					\sbra{
						\rbra{m_{H^\pm}^2-m_f^2-m_{f'}^2}^2-4m_f^2m_{f'}^2
					}^{1/2}
				\nn\\&\quad\times
					\sbra{
						\rbra{m_{H^\pm}^2-m_f^2-m_{f'}^2}\rbra{|y^+_f|^2+|y^+_{f'}|^2}
						-4m_fm_{f'}\re[y^+_f{}^*y^+_{f'}]
					}
		,\end{align}
where
$\tau^i_X=m_{H^0_i}^2/4m_X^2$,
$\delta_Z'=\frac{7}{12}-\frac{10}{9}\sin^2\theta_W+\frac{40}{27}\sin^4\theta_W$,
$\delta'_W=1$
and $\phi=H^0_{j\neq i}(H^\pm)$ for $V^*=Z^*(W^*)$. 
The couplings
$y^+_f$ and $y^+_{f'}$
are given by
$\mathcal{L}\supset (\bar{f}_R y^+_f f'_L+\bar{f}_L y^+_{f'} f'_R) H^+ +h.c.$. 
The functions $A(\tau)$ are given by
	%%%
		\eq{
				A_{1/2}^H(\tau)&=	2[\tau+(\tau-1)f(\tau)]\tau^{-2}
			,\\	A_{1/2}^A(\tau)&=	2\tau^{-1}f(\tau)
		,}
and where
	%%%
		\eq{
			f(\tau)=
				\left\{
				\begin{array}{ll}
					\arcsin^2\sqrt{\tau},
								&\quad\textrm{for}\quad	\tau\leq1
					,\\
					-\frac{1}{4}\sbra{\log\frac{1+\sqrt{1-\tau^{-1}}}{1-\sqrt{1-\tau^{-1}}}-i\pi}^2,
								&\quad\textrm{for}\quad	\tau>1
				.\end{array}
				\right. 
		}
In terms of $\lam(x,y)=1-2x-2y+(x-y)^2$ with $x=m_X^2/m_{H^0_i}^2$,
the function $G(x,y)$ is given by
	%%%
		\small
		\begin{align}
			G(x,y)=
				\frac{1}{12y}
				\Biggl\{
				&
					2(-1+x)^3 -9(-1+x^2)y +6(-1+x)y^2
					-3\sbra{1+(x-y)^2-2y}y\log x
				\nn \\
				&
					+6(1+x-y)y\sqrt{-\lam(x,y)}
					\sbra{
						\arctan\rbra{\frac{-1+x-y}{\sqrt{-\lam(x,y)}}}
						+\arctan\rbra{\frac{-1+x+y}{\sqrt{-\lam(x,y)}}}
					}	
				\Biggl\}
		\end{align}
		\normalsize

\bibliographystyle{apsrev4-1}
\bibliography{refs}

\end{document}